\pdfoutput=1
\documentclass[11pt,a4paper]{iopart}
\usepackage{hyperref}
\usepackage[british]{babel}
\usepackage[latin1]{inputenc}
\usepackage[T1]{fontenc}
\usepackage[final]{showkeys} % [draft/final]
\usepackage{enumerate}
\usepackage[]{cleveref}
\Crefname{equation}{Eq.}{Eqs.}
\crefformat{plural}{#2eqs.~(#1)#3}
\usepackage{float}
\usepackage{graphicx}  % got figures? uncomment this
\usepackage{multirow}

\makeatletter\g@addto@macro\bfseries{\boldmath}\makeatother%

\begin{document}

% title page
%\preprint{\today}

\title{Using storytelling to foster the teaching and learning of gravitational waves physics at high-school }

%\author[1,2]{Matteo Tuveri}\emailAdd{matteo.tuveri@ca.infn.it}
%\author[3]{Daniela Fadda}\emailAdd{daniela.fadda80@unica.it}

%\affiliation[1]{Dipartimento di Fisica, Universit\`a di Cagliari,\\Cittadella Universitaria, 09042 Monserrato, Italy}
%\affiliation[2]{INFN, Sezione di Cagliari}
%\affiliation[3]{Dipartimento di Psicologia, Pedagogia e Filosofia, Universit\`a di Cagliari - Cagliari, Italy}

\author{M Tuveri$^1$$^3$, A Steri$^2$$^3$ and D Fadda$^4$}

\address{$^1$ Dipartimento di Fisica, Universit\`a di Cagliari, Cittadella Universitaria, 09042 Monserrato, Italy}
\address{$^2$ Dipartimento di Ingegneria Meccanica, Chimica e dei Materiali, Universit\`a di Cagliari, 09123 Cagliari, Italy}
\address{$^3$ Istituto Nazionale di Fisica Nucleare, Sezione di Cagliari, Cittadella Universitaria, 09042 Monserrato, Italy}
\address{$^4$ Dipartimento di Pedagogia, Psicologia, Filosofia, Universit\`a di Cagliari, 09123 Cagliari, Italy}

\ead{matteo.tuveri@ca.infn.it}

\maketitle

%\listoftables

\begin{abstract}
Studies in Physics Education Research show that interdisciplinary approaches in education foster students' motivation, creativity, curiosity, and interest in physics. We discuss their features and potential role in bringing contemporary physics topics to high school, and how to use them to integrate formal educational programs. We make an explicit example of the use of storytelling and theatrical techniques to introduce secondary school students to black holes and gravitational waves topics. The activity has been designed by the Educational Division of the Physics Department at the University of Cagliari. Participants were 200 high-school students (17 to 19 years old) from five schools (scientific, humanities) in Sardinia. A measure of the efficacy in the use of artistic tools to communicate and teach the proposed subjects has been done utilizing a research questionnaire. We collected 76 answers. Results show that our methodology is useful to introduce students to contemporary physics themes, fostering their interest and learning of such contents. Students from humanities significantly appreciated more the use of poetry and artistic tools than their scientific peers. Finally, we discuss the potentiality of our approach in orientating students towards a STEAM (STEM and Arts) career. 

\end{abstract}

%%%%%%%%%%%%%%%%%%%%%%%%%%%%%%%%%%%%%%%
%%%%%%%%%%%%%%%%%%%%%%%%%%%%%%%%%%%%%%%
%\verb"\noindent{\it Keywords\/}: informal learning; interdisciplinary; storytelling; general relativity; gravitational waves

\submitto{\PED}

\section{Introduction} 
Offering an interdisciplinary vision of science to high school students and teachers is becoming a common trend of learning (see~\cite{ref:Spelt2009,ref:Gao2020,ref:Davies2007} and refs therein). Interdisciplinary, in its most simple form, constitutes an appreciation of different discipline fields and the value of communicating across disciplinary boundaries to find ways to work together. As emphasized in~\cite{ref:ThongBook2023}, playing and learning contents and methods from different disciplines bring educators and students to develop new forms of scientific reasoning and toolkits.
Learning experiences and skills acquired through interdisciplinary apply physics fundamentals to real-world solutions making a connection between the world of research and society. 
To show how science is evolving and to provide new instruments to learn science and physics in an enlarged context, mixing knowledge, techniques, and methods from different disciplines should be part of science and education curricula, developing an integrated model of learning and teaching~\cite{ref:Khalick2010}.
Interdisciplinary approaches in education also enhance students' engagement in physics~\cite{ref:GilibertiBook2023}.    

In general, this kind of educational experiments are carried out in informal context, such as out-of-school time, and outdoor activities, science outreach labs, visits to museums, summer camps, where they keep in contact with science through different media and activities~\cite{ref:Fazio2021,ref:NRC2009,ref:Ucko2010,ref:Tuveri2023_Udine}. Researches show that teaching science in informal contexts increases experience excitement, interest, and motivation to learn about phenomena in the natural and physical world, improves scientific awareness and literacy~\cite{ref:Giliberti2019,ref:Giliberti2022,ref:Michelini2021,ref:Corni2021,ref:Michelini2010}, bolstering motivation to learn science and physics~\cite{ref:Michelini2005,ref:Affeldt2017,ref:Sokolowska2018}. It also helps them in building a \lq\lq culturo-scientific thinking\rq\rq, that is learning how to think scientifically and creatively, with an understanding of the evolution of the nature of science~\cite{ref:Goorney2022}. 

Integrating the good practices of learning in informal contexts even in formal curricula at school could bring some benefits in students' learning of science, engagement and motivation.  
Mixing arts and science seem to be very promising in this direction~\cite{ref:GilibertiBook2023,ref:Fazio2021}. Research results show that arts, narrative, and drama help students in reflecting on scientific concepts and science evolution from epistemological and historical perspectives. It promotes peer interaction at school and interaction among school, family, and society~\cite{ref:Odegaard2003}, also preventing students from school abandon~\cite{ref:Scierri2019}. Moreover, it helps teachers better understand what students are thinking~\cite{ref:McGregor2014}. This kind of approach involves emotional involvement to foster students's development of scientific imagination and creativity, allowing personal learning styles, and cognitively mediating in approaching physics (see~\cite{ref:GilibertiBook2023} and refs therein). Moreover, when followed by minds-on activities, such as a debate, students re-elaborate what they experienced and/or re-propose contents in a personal way, bettering their learning process~\cite{ref:Giliberti2019,ref:Giliberti2022}, developing their critical thinking skills. These activities also orientate students towards a STEM career~\cite{ref:Nguyen2018,ref:Devins2015}.

To show how science is evolving and to provide new instruments to learn science and physics in an enlarged context, mixing knowledge, techniques, and methods from different disciplines should be part of science and education curricula, developing an interdisciplinary and integrated model of learning and teaching~\cite{ref:Spelt2009,ref:Gao2020,ref:Davies2007,ref:Khalick2010,ref:TuveriBook2023}.
Nevertheless, sometimes secondary science teachers face some difficulties in inspiring their classes. The use of stories can help improve teaching and students' learning and physics is full of stories that can afford for this job. The use of narrative and stories to teach physics has some well-recognized benefits in education, presenting science in a more \lq\lq coherent, memorable, and meaningful format in order to interest and engage pupils, as well as to unify the curriculum and to preserve a sense of the bigger picture\rq\rq~(see~\cite{ref:Rowcliffe2004}, p.121). Storytelling can be used as a stimulus to engage, excite, and emotionally involve experience from which to learn, also bettering students' capability to memorize concepts~\cite{ref:Rowcliffe2004,ref:Rawatee2022}. Millar and Osborne~\cite{ref:Millar1998} enlightened the role of sustaining and developing the curiosity of young people attending a science curriculum, fostering a sense of wonder, enthusiasm, and interest in science. It seems that stories could be a way of achieving this sense of wonderment, enthusiasm, and interest in science, provided they are relevant, fun, and interesting~\cite{ref:Rowcliffe2004}.

Despite its role in education is widely recognized in many contexts and, in particular, in science education~\cite{ref:Rawatee2022,ref:Engel2018,ref:Kaur2020}, the use of storytelling in school is not so common. Since storytelling has its genesis in arts, someone speculates that is not recognized as a rigorous pedagogical methodology to teach and learn~\cite{ref:Denning2005}. This idea can be mediated by the fact that instructional approaches that are perceived to be subjective and less rigorous will only be adopted with cautious reservation, basically for science communication~\cite{ref:Richter2019}, not for science lessons~\cite{ref:Olson2015}. However, as noted in~\cite{ref:Petrucco2009}, teaching itself is, in reality, nothing more than an evolved and extremely codified form of storytelling, the more effective the more it is linked to the telling of real stories that highlight from time to time crucial elements of the discipline being learned.

Studies in Physics Education Research (PER) underly that storytelling fosters pupils' creativity and positive attitude towards science~\cite{ref:Simon2000}. It also offers a methodology guide to teach science in the laboratory. For example, it can be used to introduce a scientific problem in the form
 of a story for students to solve, possibly incorporating real-life scientific issues that researchers face in their everyday life~\cite{ref:Rowcliffe2004,ref:Richter2019}. 
Storytelling has always been a medium to transfer traditional knowledge, beliefs, values, and practices over time and its core relies on sound education pedagogy~\cite{ref:Abrahamson1998,ref:Farrel1982,ref:McDonald2009}. Formal programs in school included aspects of storytelling that were used to raise
students' interest and to encourage students to actively participate in learning~\cite{ref:Green2002}, but recently new forms of digital storytelling seem to replace them~\cite{ref:Petrucco2009,ref:Kotluk2016,ref:Marsico2019}.
Nevertheless, role-play, dramatization, and storytelling will continue to hold promise for the creation of
exciting, engaging, meaningful, and rewarding classroom experiences even in a technological environment storytelling will remain a versatile option available to teachers to introduce science in the classroom.

The specific focus of this work is on the use of storytelling to bring contemporary physics topics in high school and, in particular, black hole and gravitational waves physics. Specifically, the aim of this work is to explore the effect of storytelling on students' levels of classroom participation, motivation, and interest in the proposed topics. We also meant to measure their engagement and, most interestingly, their views about the effectiveness of storytelling as a teaching/learning strategy in the science classroom. The research can give instructors a methodological tool to encourage them to bring these topics to school, using storytelling to optimum advantage in science. This introduces students to current trends in research, trying to bypass content-related difficulties (both physical and mathematical), but still making them explore our universe with inquiry and minds-on activities, improving their motivation, curiosity, and interest in physics.

In this paper, we address our research goal by reporting from an educational program developed by the PER group at the University of Cagliari which makes use of storytelling and theatrical techniques (a monologue) to bring general relativity and gravitational-related topics such as gravitational waves and black hole physics at high-school. We briefly illustrate the physical background and the design of the educational program. The activity involved 200 students (17 - 19 years old) from five schools in Sardinia. We measured the efficacy of our activity according to our research goals by using a questionnaire. We report and discuss our results, pointing the potentiality of our approach out in orientating students towards a STEAM (STEM and arts) career.

%%%%%%%%%%%%%%%%%%%%%%%%%
\section{Physics Framework and the monologue}
In the context of contemporary physics, General Relativity (GR) gives the possibility to carry out some educational experimentation taking into account the use of stories and storytelling. Indeed, in a broader sense, GR topics are related to the world views held by students, such as our universe, stars, and planets, thus providing conceptual, theoretical, and experimental means for bridging the gap between the teaching of science and the teaching of humanistic subjects. This is true, for example, for concepts pertaining to cosmology like the Big Bang, which connects physics with philosophy~\cite{ref:Kragh2013,ref:Kragh2011}. Students' first approach to these contents is different: movies, documentaries, media, and TV shows, all of which contribute to shaping their conceptions about astrophysical phenomena. The work of experts should be that of guiding them toward the understanding of our universe, stimulating the debate, and offering a modern vision of phenomena around us~\cite{ref:Fazio2021,ref:Giliberti2022}. This also offers the possibility to debate about what is science and what is not, what can be measurable and what is not, thus re-enforcing their critical thinking~\cite{ref:Kragh2011}. 

Let us make an explicit example for the sake of completeness. The concepts of \lq\lq Big Bang\rq\rq~and \lq\lq black hole\rq\rq~are metaphors, analogies used by physicists to explain something invisible whose physical meaning is still obscure, but which is useful to imagine a given phenomenon. They are a linguistic expedient that represents a way to move from a provisional ad hoc model (for example an explosion or a dark, deep hole, respectively) to a permanent theory (GR)~\cite{ref:GilibertiBook2023,ref:Ogborn2011,ref:TuveriBook2023}. From the educational point of view, they have the potential to grasp and hold students' attention while easily encouraging them to participate, share, and collaborate to learn. Telling the universe by means of such metaphors is not only an artistic and linguistic tool, but it has also an educational purpose, giving the possibility to teachers to explore contemporary physics topics in class, living the contemporaneity of research and its methodologies. 

More in detail, the word \lq\lq black hole\rq\rq, invented by John A. Wheeler~\cite{ref:Thorne2009}, describes the final state of matter when a massive star (with more than 1.4 Solar masses) explodes. Certainly, there are not any \lq\lq holes\rq\rq~in our universe, but the idea of an obscure region of spacetime where everything falls down and objects cannot escape from there is pictorially represented by that metaphor. Nevertheless, even if we cannot observe them, we can study them by the motion of stars and matter surrounding them or, by detecting the gravitational waves produced by the merging of two black holes. Indeed, in 2016, after 100 years of Einstein paper~\cite{ref:Einstein1916}, the LIGO and Virgo collaborations first announced the discovery of gravitational waves, ripples of space-time traveling all around the universe at the speed of light produced by violent astrophysical events~\cite{ref:Abbott2016}. An interesting feature of such waves is that their effect is to modify space and time, measuring changes in length much smaller than the diameter of an atomic nucleus (for a merging of black holes with masses of the order of ten Solar masses, the change in length is of the order of $10^{-21}$ cm, see~\cite{ref:Abbott2016}).  
Moreover, in 2019, researchers from the Event Horizon Telescope collaboration published a \lq\lq picture\rq\rq~of the first image of the shadow of the supermassive black hole at the center of the M87 galaxy~\cite{ref:EHT2019}, replicated in 2022 with the first image of the black hole at the center of our galaxy, the Milky Way~\cite{ref:EHT2022}. 

The search for understanding our universe is still ongoing and researchers are looking for new technologies and devices to study its features, mainly in its initial stages. Indeed, when stars were not formed at all, no light signal could be detected from satellites or telescopes on the ground, so standard astronomical observation cannot give us any information about the first million years of our universe. At that time, no matter or light signal was there and, basically, only primordial quantum fluctuations populated the universe, probably growing, collapsing, and forming primordial black holes. Thus some primordial gravitational signal could, in principle, be detectable~\cite{ref:Maggiore2020,ref:Branchesi2023}.
This is, for example, one of the goals of the Einstein Telescope (ET) experiment~\cite{ref:ET2010}, a ground-based third-generation gravitational wave interferometer aimed to explore the universe with gravitational waves up to cosmological distances. ET has a reference configuration based on a triangular shape consisting of three nested detectors with 10 km arms, where each detector has a \lq\lq xylophone \rq\rq~configuration made of an interferometer tuned toward high frequencies, and an interferometer tuned toward low frequencies and working at cryogenic temperature~\cite{ref:Branchesi2023}. Its scientific output is related to compact binary coalescences, multi-messenger astronomy, and stochastic backgrounds.  The detector will be capable of observing the entire Universe using gravitational waves aiming to increase a factor ten the sensitivity of previous generation detectors~\cite{ref:Grado2023}. It could be operating in the mid-2030s in Europe, and one of the site candidates is in Sardinia (Italy). This site has peculiar features (very low anthropomorphic, electromagnetic, and geophysical noises in the range of frequencies at which ET will operate) which makes it one of the best candidates to host the infrastructure~\cite{ref:Saccorotti2023}.

\subsection{The monologue}
The possibility for Sardinia to host the experiment motivates us to design a suitable activity to bring gravitational waves physics contents and ET physics to high school. From a methodological point of view, we use storytelling and theatrical techniques such as a monologue. The latter is called \lq\lq Viaggio in una storia lunga 14 miliardi di anni\rq\rq~(\lq\lq A 14 billions years old journey\rq\rq). 
The story time is set in the middle of 2030 when ET should be operating. The protagonist of the story is a physicist who is currently part of the ET collaboration, working on gravitational wave physics. He is in the control room of ET in Sardinia, checking for all the parameters coming from the detector in the underground caves, from optics to lasers. While he is thinking at his past, his career and his life as a researcher, his studies, and his work in Sardinia, he is also checking for possible gravitational waves signals appearing on a monitor in front of him. It was at that moment that he started traveling all around the universe, from Sardinia to the origin of the universe. In this 14 billion years journey back in time, he describes the relativistic effects due to traveling back in time at the speed of light, the structures in our universe and their formation, unless he encounters a mysterious phenomenon he realizes to be a black hole merging. This offers him the possibility to tell the physics of tidal disruption and the consequent formation and propagation of gravitational waves. Finally, he travels back home surfing such gravitational wave. Once went back to the control room, he realizes having discovered a primordial gravitational wave signal. 

The monologue (in Italian) can be found at the following link:~\url{https://www.youtube.com/watch?v=SMaNFRKdy4c}.

\begin{figure}[h]
\centering%
\includegraphics[width=0.5\textwidth]{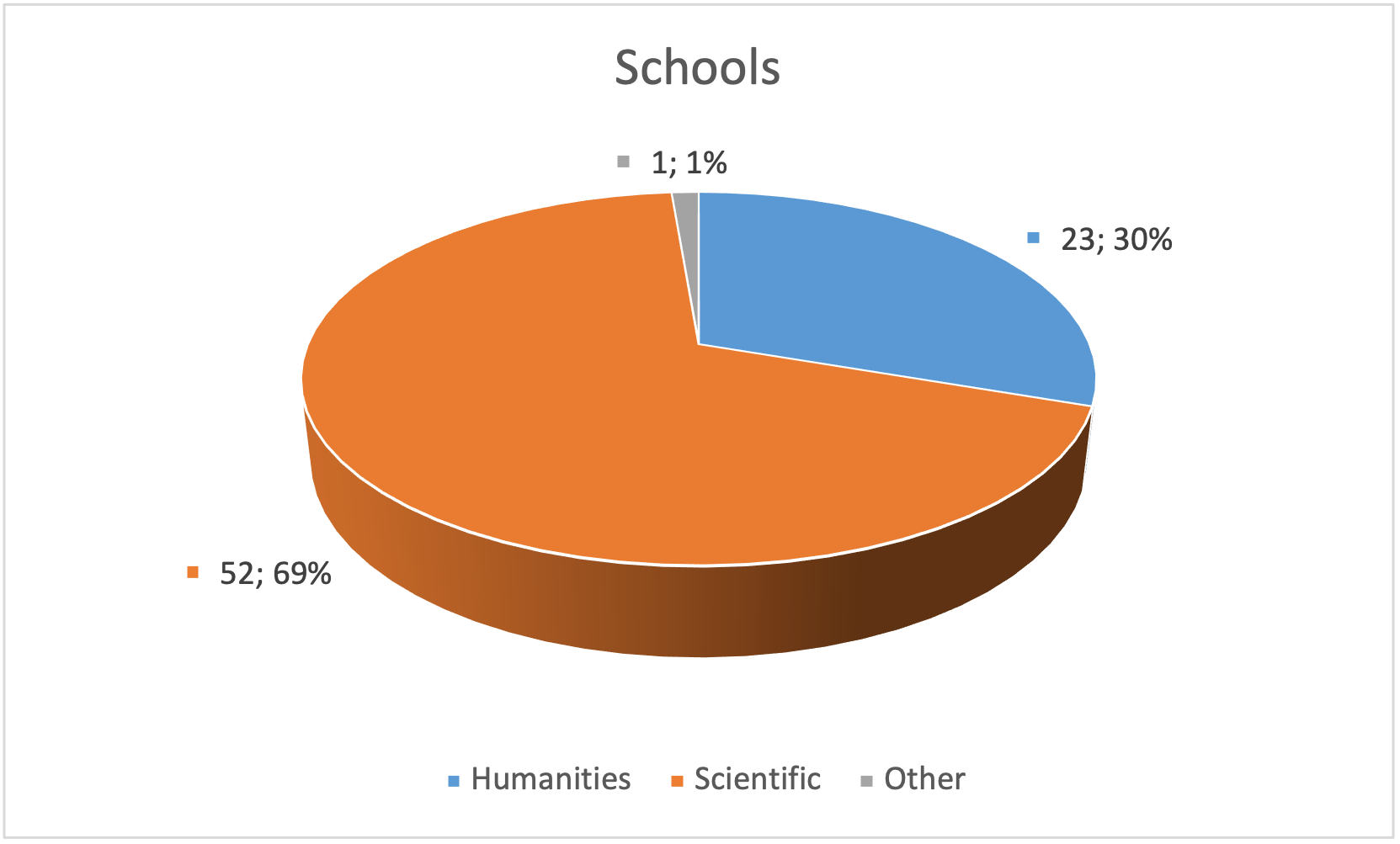}  
\caption{The panel shows students' distribution as a function of school.}\label{fig:ETschool}
\end{figure}

\begin{figure}[htbp]
  \centering
  \begin{minipage}[b]{0.45\textwidth}
    \includegraphics[width=\textwidth]{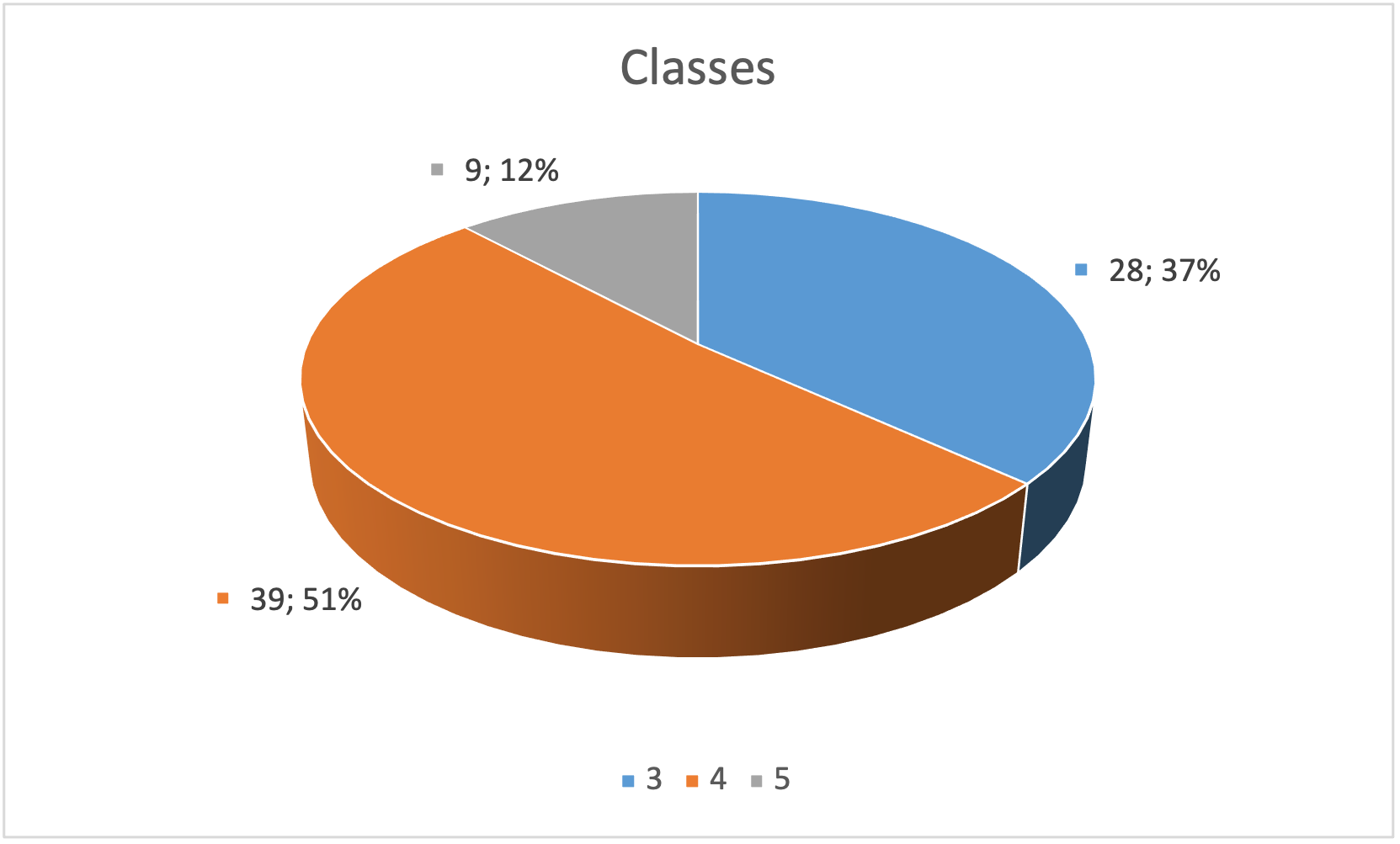}
    %\caption{The panel shows students' distribution as a function of the classes.}
  \end{minipage}
  \hfill
  \begin{minipage}[b]{0.45\textwidth}
    \includegraphics[width=\textwidth]{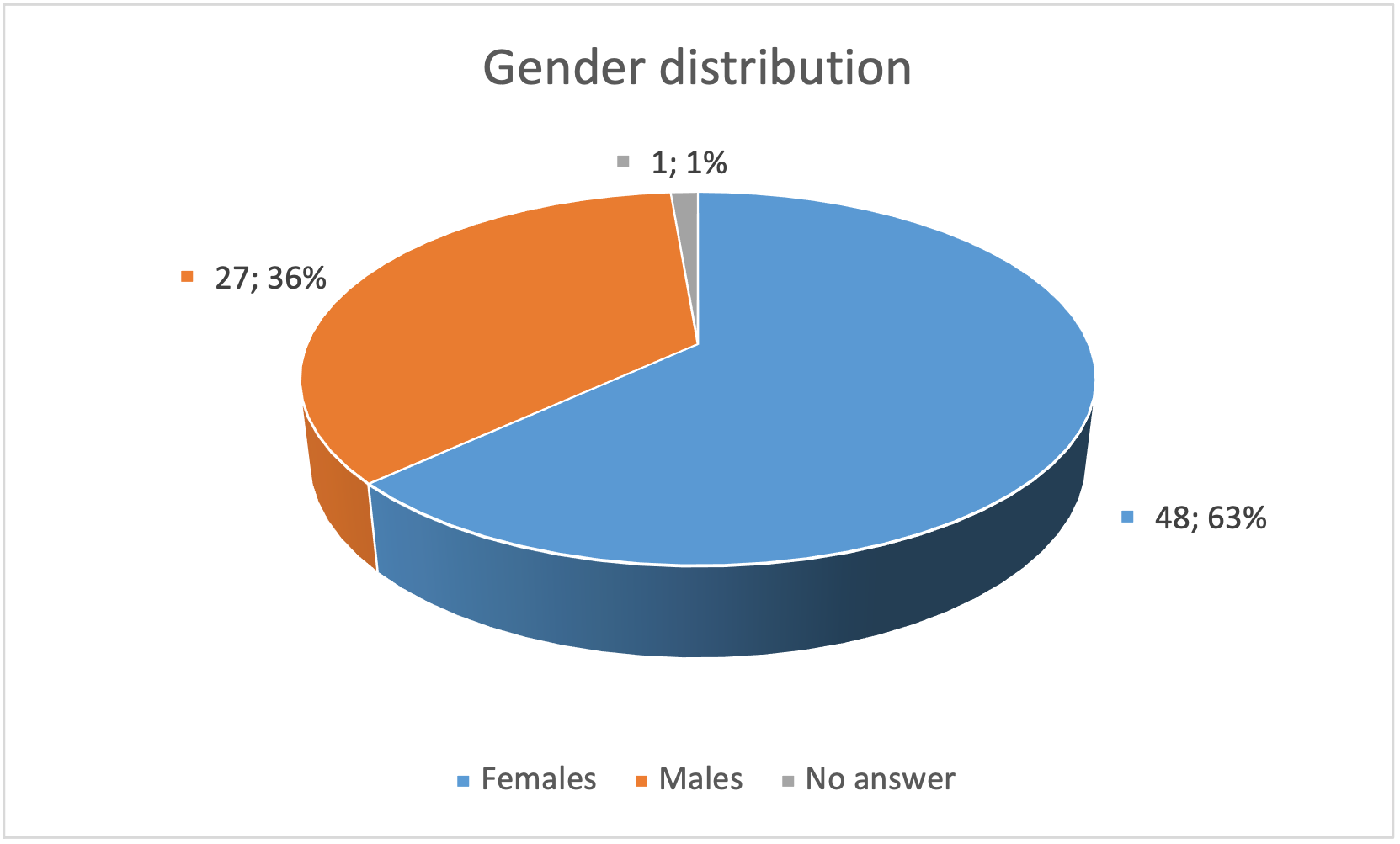}
    %\caption{The panel shows students' distribution as a function of the gender.}
  \end{minipage}
  \caption{The panel shows students' distribution as a function of the classes (left) and the gender (right).}\label{fig:ET_class_gender}
\end{figure}

%%%%%%%%%%%%%%%%%%%%%%%%%
\section{Methods}
The activity was held in Sardinia, Italy, in 2023 along the lines of the informal learning educational program at the Physics Department of the University of Cagliari and the Cagliari Division of the National Institute of Nuclear Physics (INFN). The activity involved 200 students (17-19 years old) from five high schools in all the Region (4 scientific lyceums, 1 humanities). In Italy high school lasts five years, students participating to the activity were attending third to fifth year classes. 

The activity was organized as follows: the first 20 minutes was dedicated to the monologue, followed by a 20-minute session where the a researcher explained gravitational-related contents and ET physics (gravity according to general relativity, black holes, gravitation waves formation and propagation, detection of gravitational waves with ET, cosmology). Finally, 20 minutes were left for debate, discussions, and a question and answers section. 

During the monologue, some poems both in Italian and Sardinian language~\footnote{The Sardinian language, with its several variants all along the Island, descends from Latin with some words remaining from the ancient Nuragic language, and it is currently spoken in Sardinia. It is generally non longer taught in school, a part from some specific project pertaining mostly primary schools, but people learn it mostly at home.} were used to tell some facts concerning physics. This served as an artistic tool and also to reinforce the identity of the character which is also related to the place where ET is located in the story.   

To address the aim of this work, which is to explore the effect of storytelling on students' levels of classroom participation, motivation, and interest in the proposed topics, we built a specific research questionnaire. In this way, we could quantitatively measure their engagement and their perception about the efficacy of storytelling as a teaching/learning strategy in the science classroom. We identified five dimensions corresponding to eight aspects to investigate: general information (demography - 5 items; preliminary knowledge on the proposed topic - 6 items); interest in STEM and engagement (interest and passion in STEM - 2 items; engagement - 4 items); communication and artistic tools (the monologue - 8 items; the poems - 5 items); motivation (3 items); final remarks (8). 

Students could answer using a 6-point Likert scale from 1 (completely disagree) to 6 (completely agree). We conducted descriptive analysis of data. After that, to determine whether there were statistically significant differences between means of domains based on schools, classes, and gender we carried out a multivariate analysis of variance (MANOVA). We also tested correlations among the domains with the \lq\lq r\rq\rq~of Pearson coefficient. Reliability of the questionnaire were investigated in terms of internal consistency by Cronbach's alpha; mean value among the domains was 0.89 (higher than the standard 0.8 threshold).

%%%%%%%%%%%%%%%%%%%%%%%%%%%%%%%%%%%%%%%
%%%%%%%%%%%%%%%%%%%%%%%%%%%%%%%%%%%%%%%
\section{Results}
We collected 76 answers, 23 from humanities and 52 from scientific high schools students (1 missing answer), see Fig.~\ref{fig:ETschool}. Considering the gender, 48 were females, 27 males, and 1 \lq\lq other\rq\rq. Concerning classes, 28 were attending the third class, 39 the fourth, and 9 the fifth class, see Fig.~\ref{fig:ET_class_gender}.  We first show general results, then we will display school, class, and gender results. After a general overview, we will focus on school, class, and gender results by 75 students (removing the case with missing data; i.e., \lq\lq other\rq\rq).

Firstly, we asked students if before attending the activity they knew the ET project: 45 answered \lq\lq no\rq\rq , 31 \lq\lq yes\rq\rq. 

\begin{figure}[h]
\centering%
\includegraphics[width=0.4\textwidth]{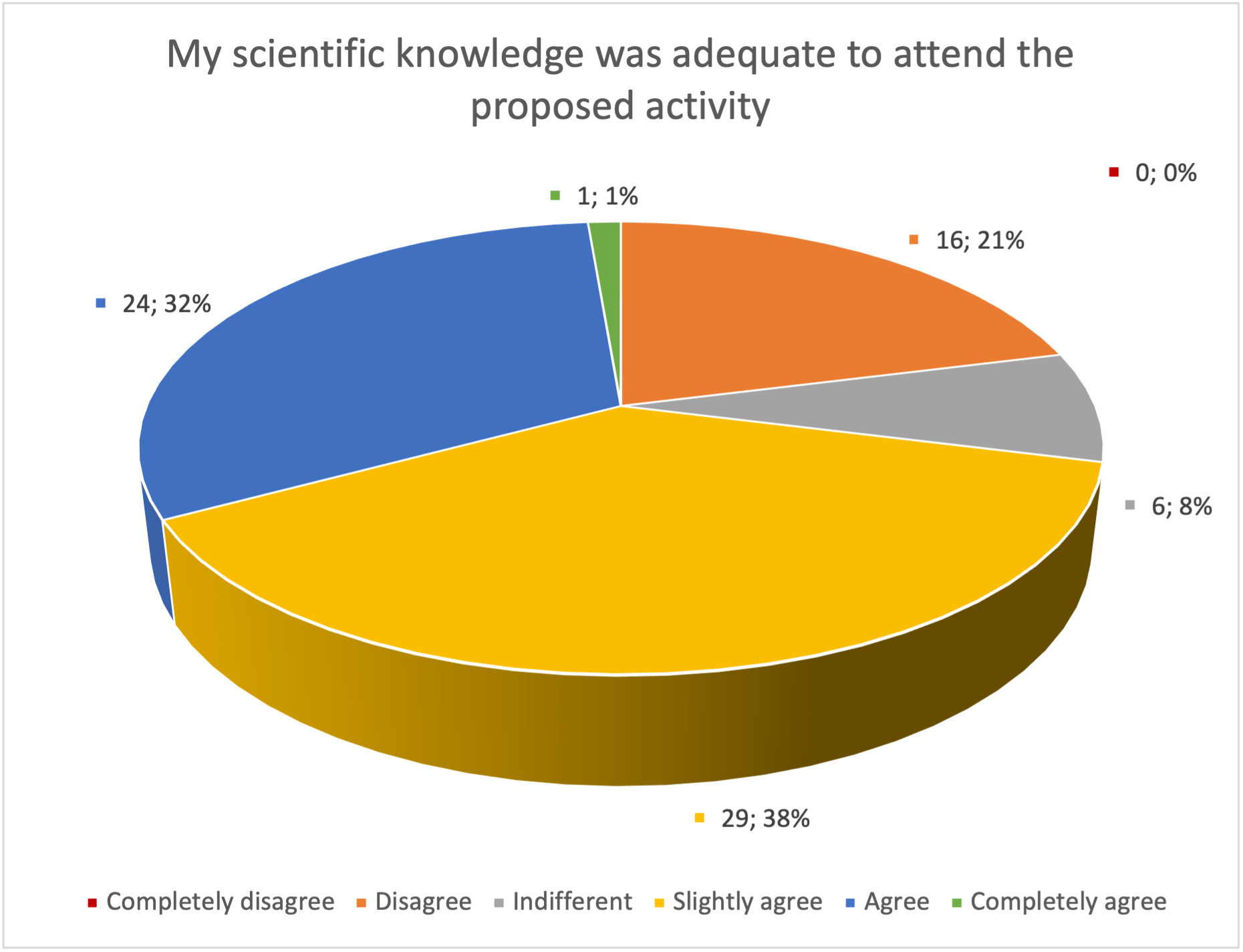}  
\caption{The panel shows students' feedback on their level of preparation to attend the activity.}\label{fig:knowledge}
\end{figure}

%general
Concerning the first macro domain, \lq\lq general information\rq\rq, we asked if students' preliminary knowledge was adequate to attend the activity: for 16 (21\%) of them, their preparation was not adequate; 6 (8\%) did not know; 29 (38\%) answered that their preparation was slightly adequate; for 24 (32\%) students their preparation was adequate and 1 student reported to have a very adequate preparation (see Fig~\ref{fig:knowledge}).   

\begin{figure}[htbp]
  \centering
  \begin{minipage}[b]{0.45\textwidth}
    \includegraphics[width=\textwidth]{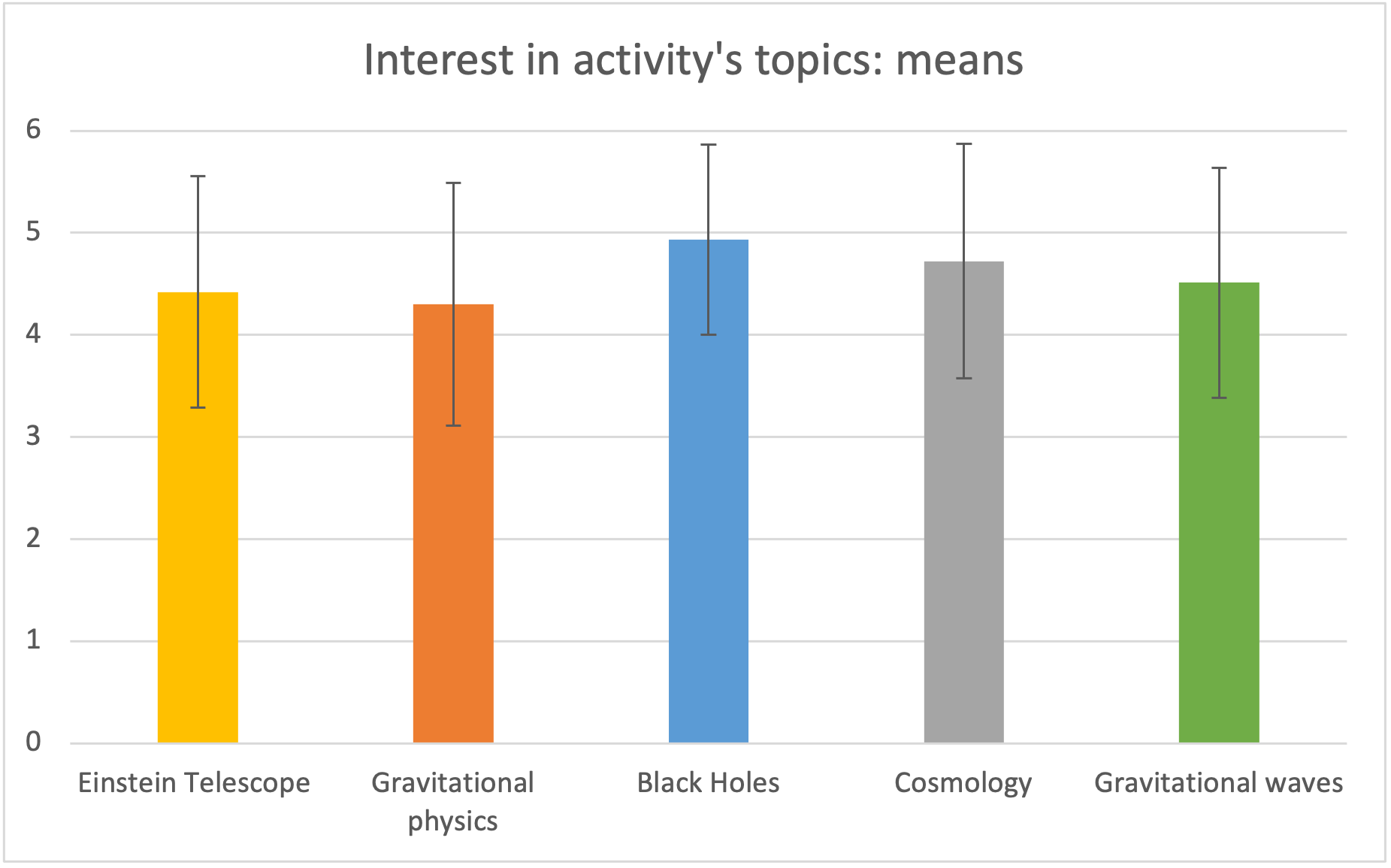}
    %\caption{The panel shows students' distribution as a function of the classes.}
  \end{minipage}
  \hfill
  \begin{minipage}[b]{0.45\textwidth}
    \includegraphics[width=\textwidth]{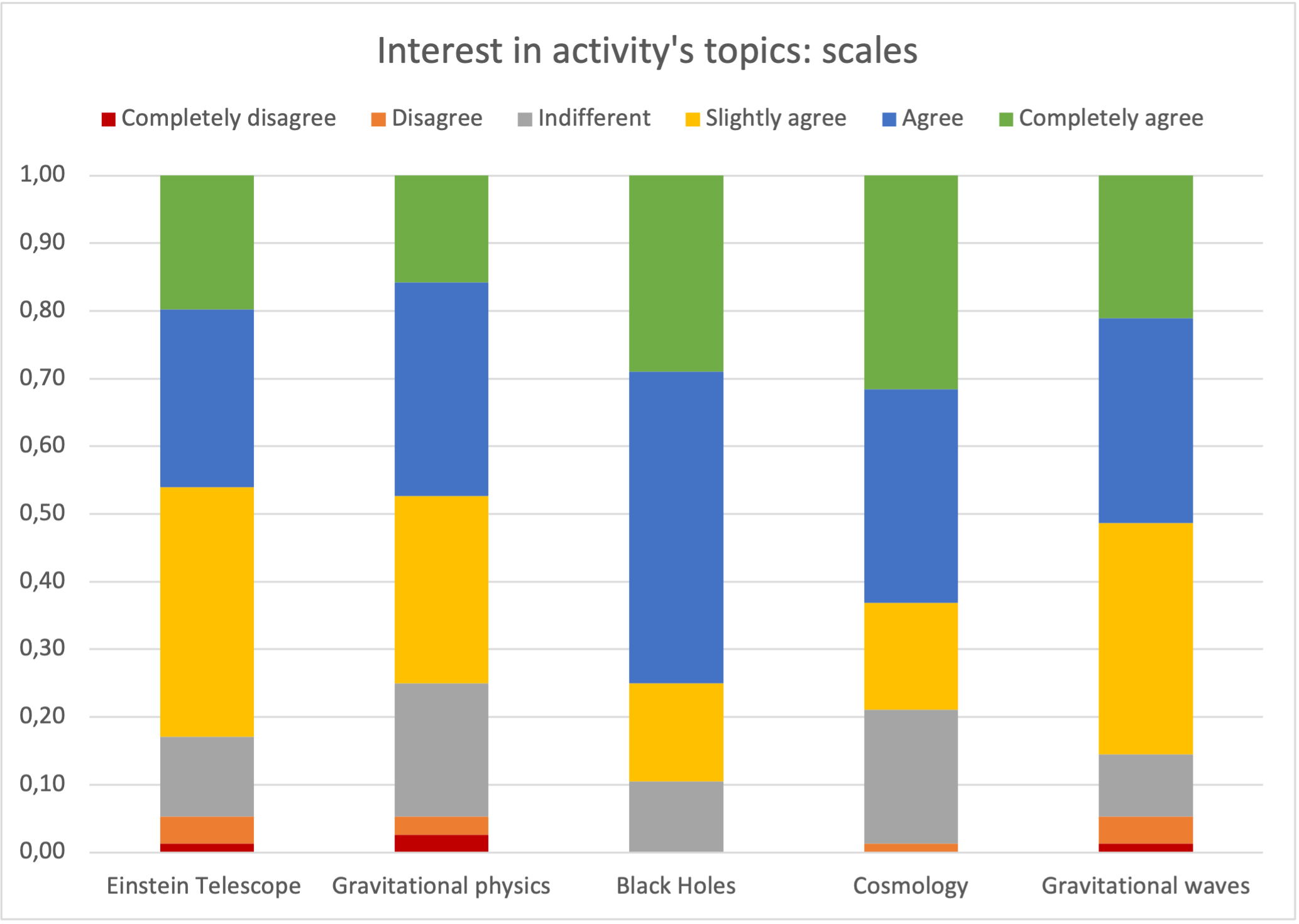}
    %\caption{The panel shows students' distribution as a function of the gender.}
  \end{minipage}
  \caption{The panel shows students' interest in topics presented during the activity. On the left, means and standard deviation; on the right, results on the 6-point Likert scale for each topic.}\label{fig:ET_interest}
\end{figure}

\begin{figure}[h]
\centering%
\includegraphics[width=0.7\textwidth]{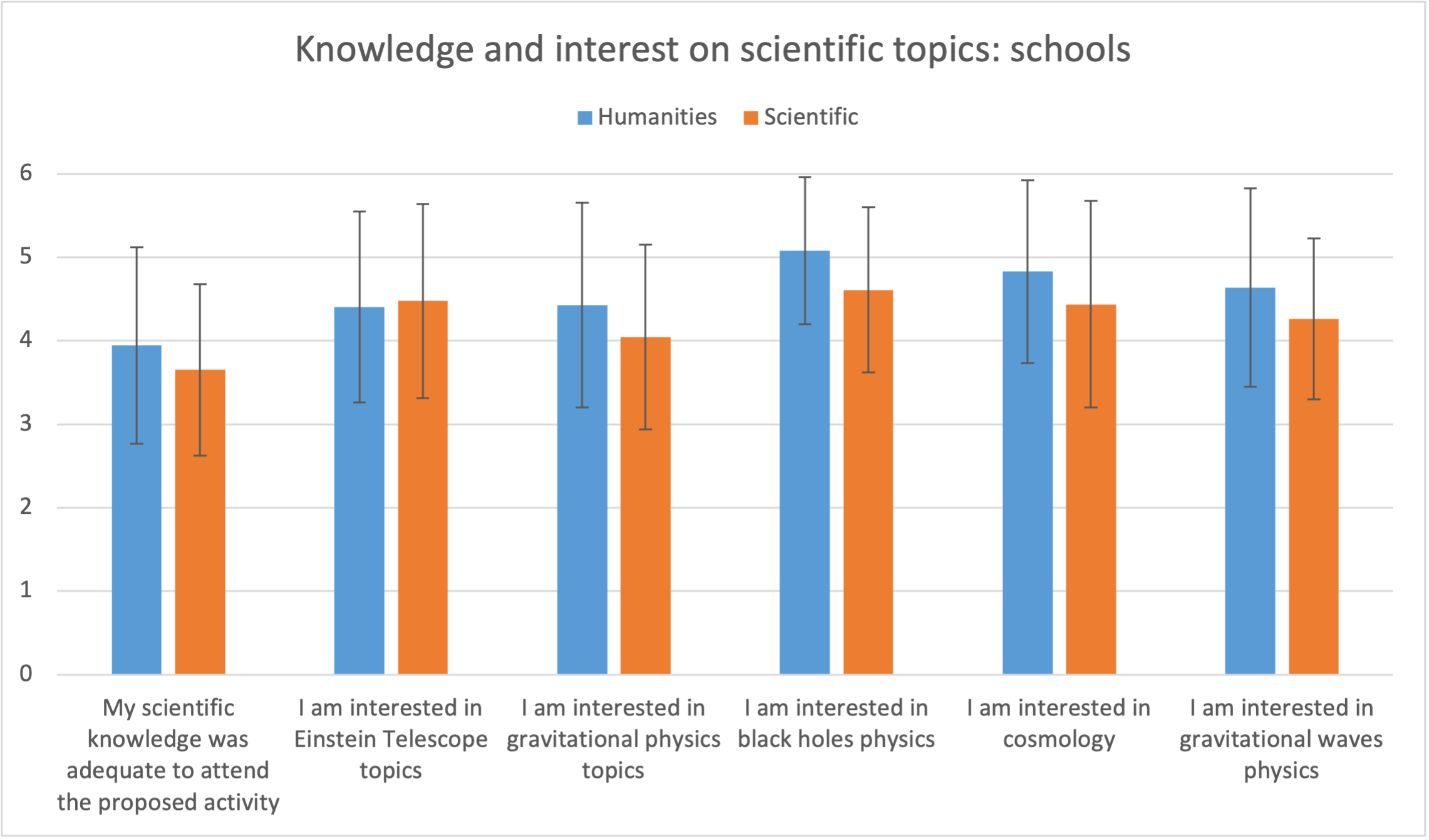}  
\caption{The panel shows means and standard deviation (error bars) about students' interest in the proposed topics according to the type of school. Students rated using a 6-point Likert scale, from 1 (completely disagree) to 6 (completely agree)}\label{fig:school_interest}
\end{figure}

\begin{figure}[h]
\centering%
\includegraphics[width=0.7\textwidth]{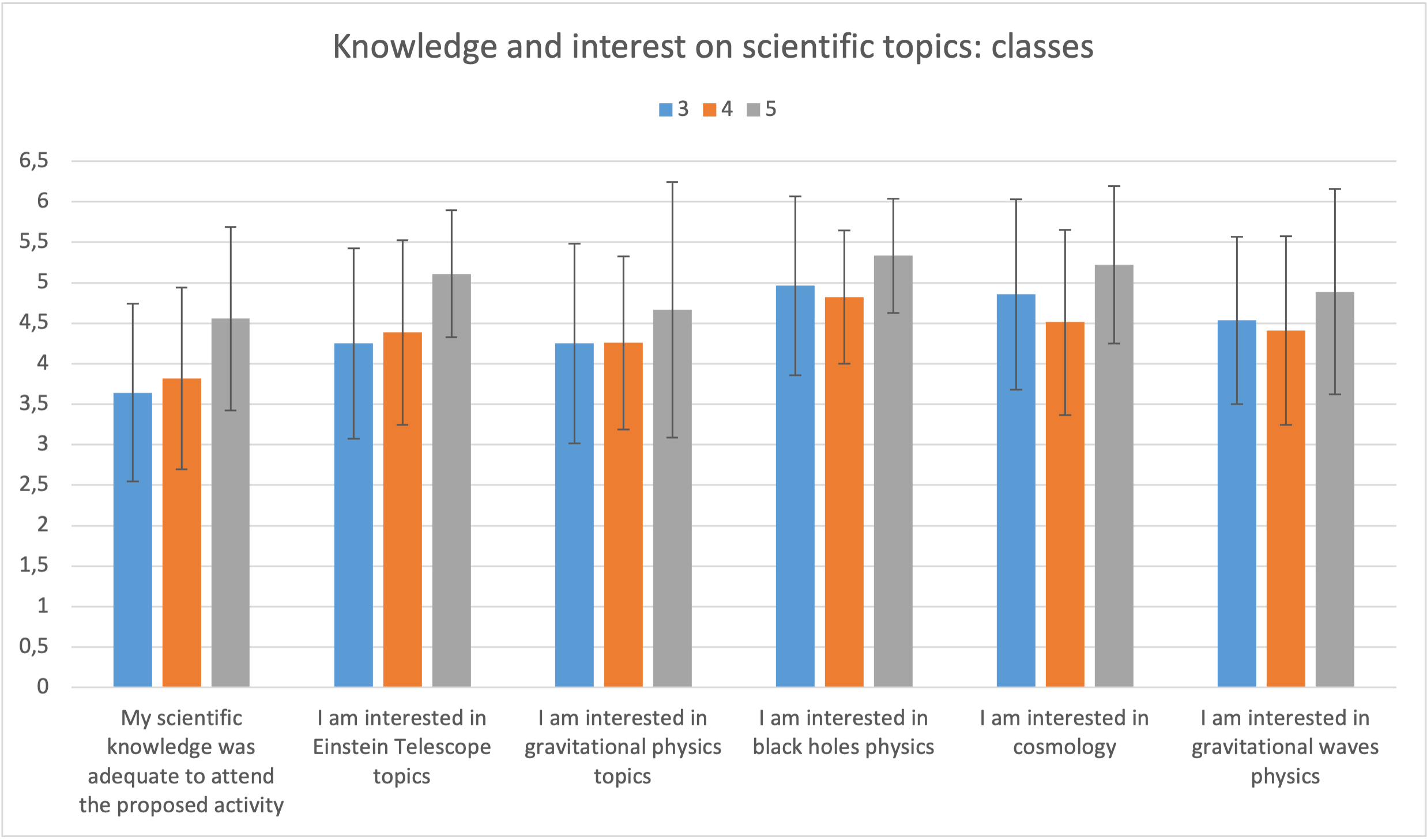}  
\caption{The panel shows means and standard deviation (error bars) about students' interest in the proposed topics according to the class. Students rated using a 6-point Likert scale, from 1 (completely disagree) to 6 (completely agree)}\label{fig:class_interest}
\end{figure}

\begin{figure}[h]
\centering%
\includegraphics[width=0.7\textwidth]{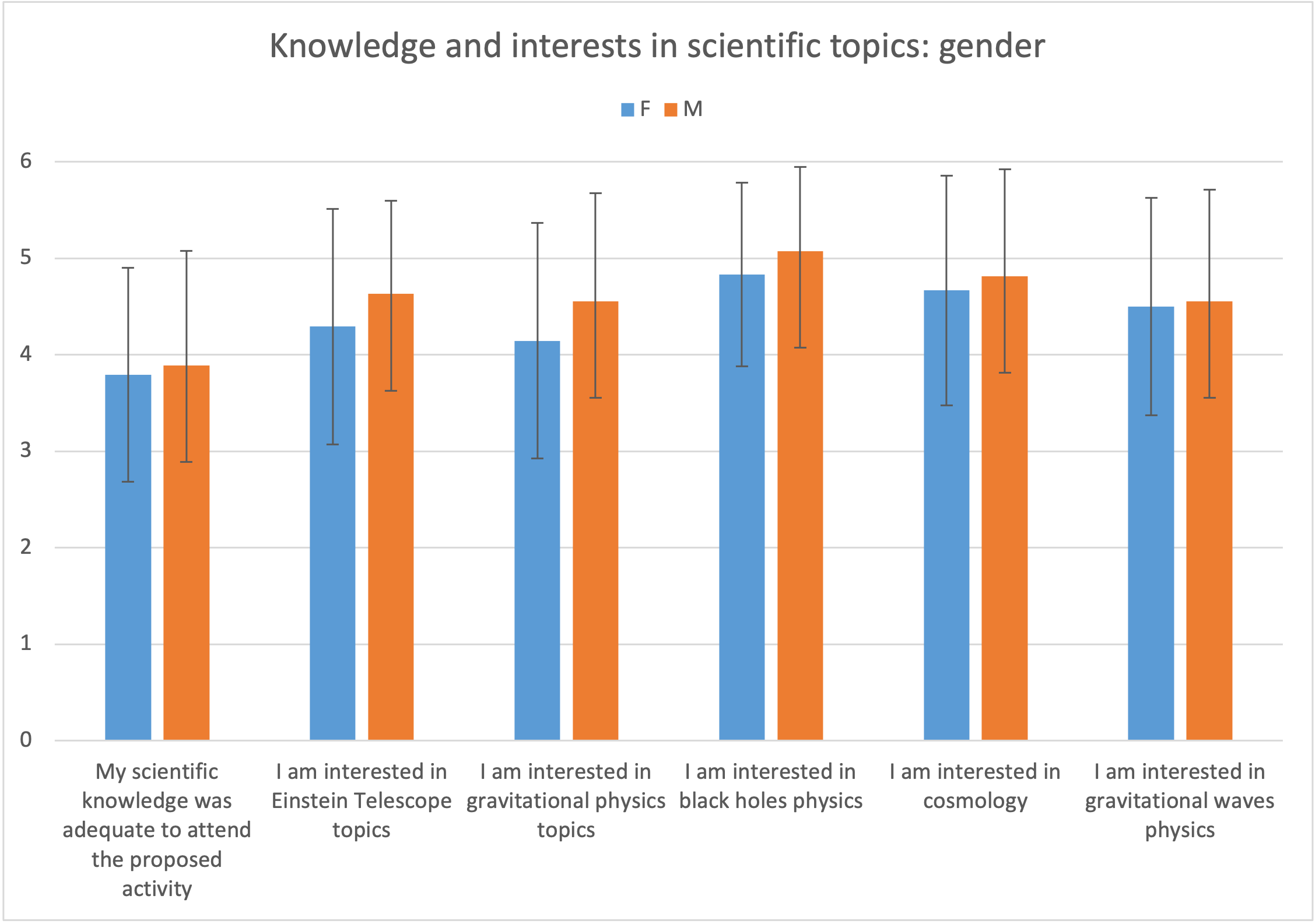}  
\caption{The panel shows means and standard deviation (error bars) about students' interest in the proposed topics according to their gender. Students rated using a 6-point Likert scale, from 1 (completely disagree) to 6 (completely agree)}\label{fig:gender_interest}
\end{figure}

We asked also for their interest in the topics we proposed. Results are shown in Fig.~\ref{fig:ET_interest}. Concerning the type of school, results are shown in Fig.~\ref{fig:school_interest}. Accordingly, results concerning the class and the gender are shown in Figs~\ref{fig:class_interest} and ~\ref{fig:gender_interest}.

%STEM ENGAGEMENT
%
\begin{figure}[h]
\centering%
\includegraphics[width=0.6\textwidth]{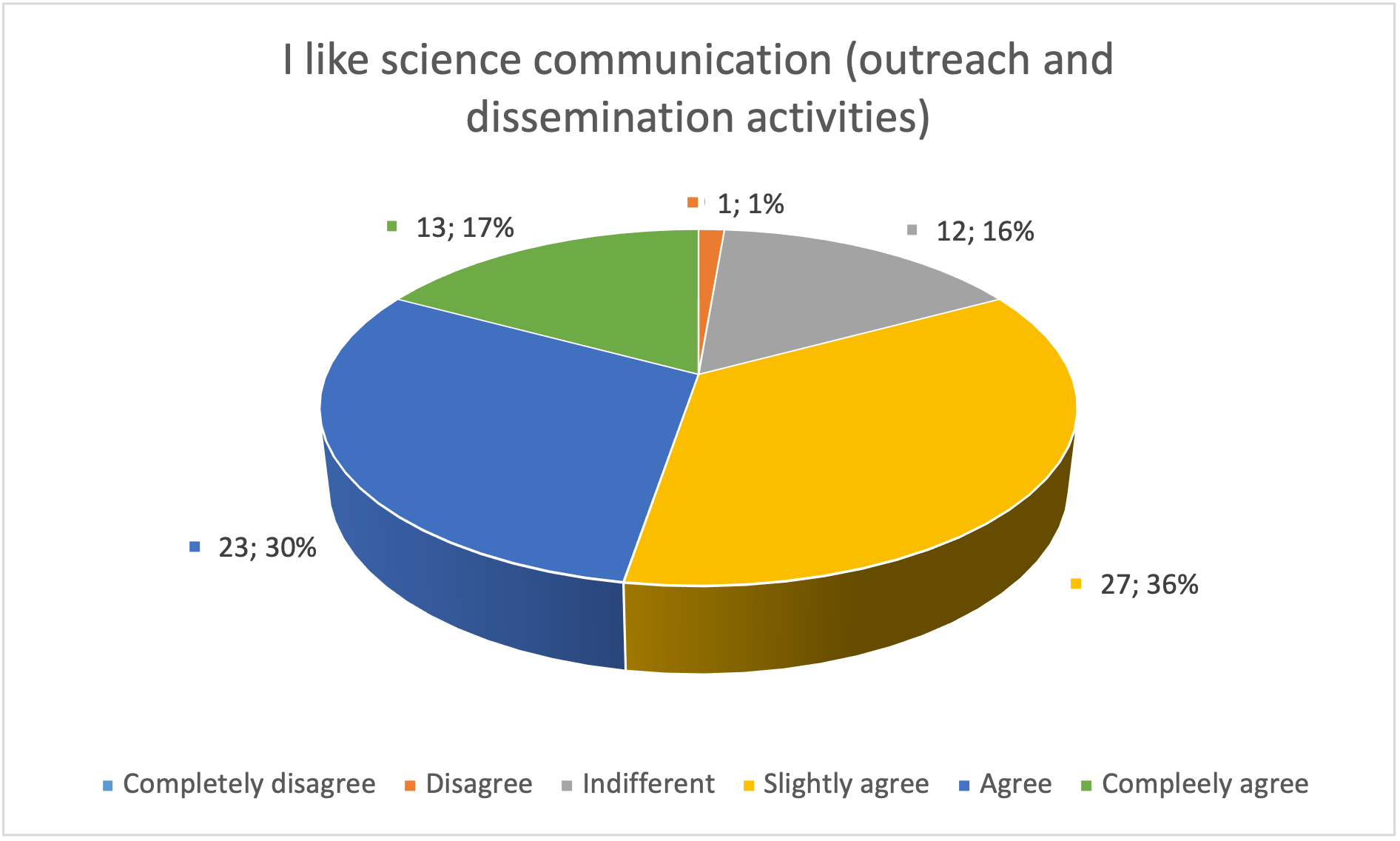}  
\caption{The panel shows students' interest in science communication. Students rated using a 6-point Likert scale, from 1 (completely disagree) to 6 (completely agree).}\label{fig:interest_outreach}
\end{figure}

\begin{figure}[htbp]
  \centering
  \begin{minipage}[b]{0.45\textwidth}
    \includegraphics[width=\textwidth]{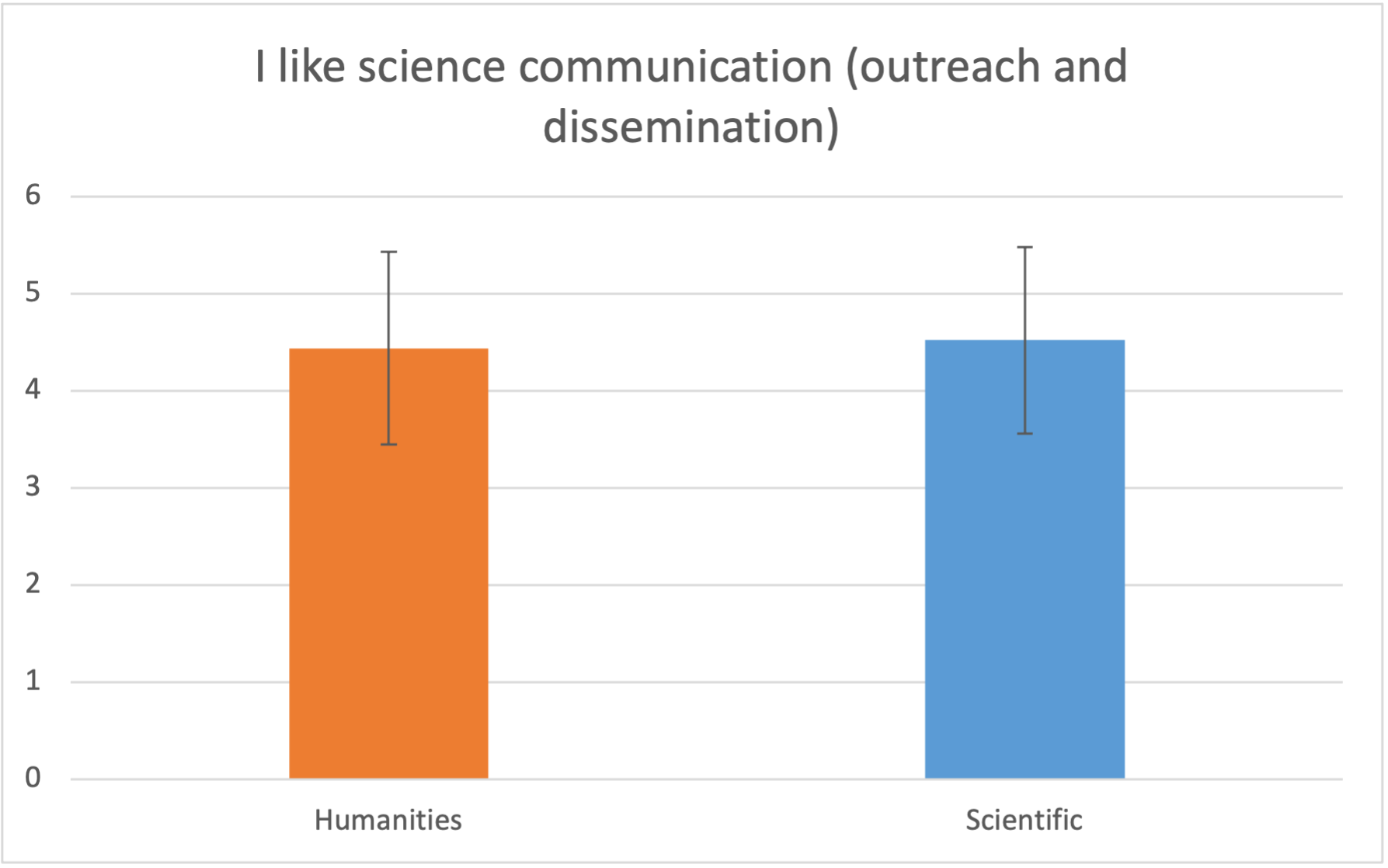}
    %\caption{The panel shows students' distribution as a function of the classes.}
  \end{minipage}
  \hfill
  \begin{minipage}[b]{0.45\textwidth}
    \includegraphics[width=\textwidth]{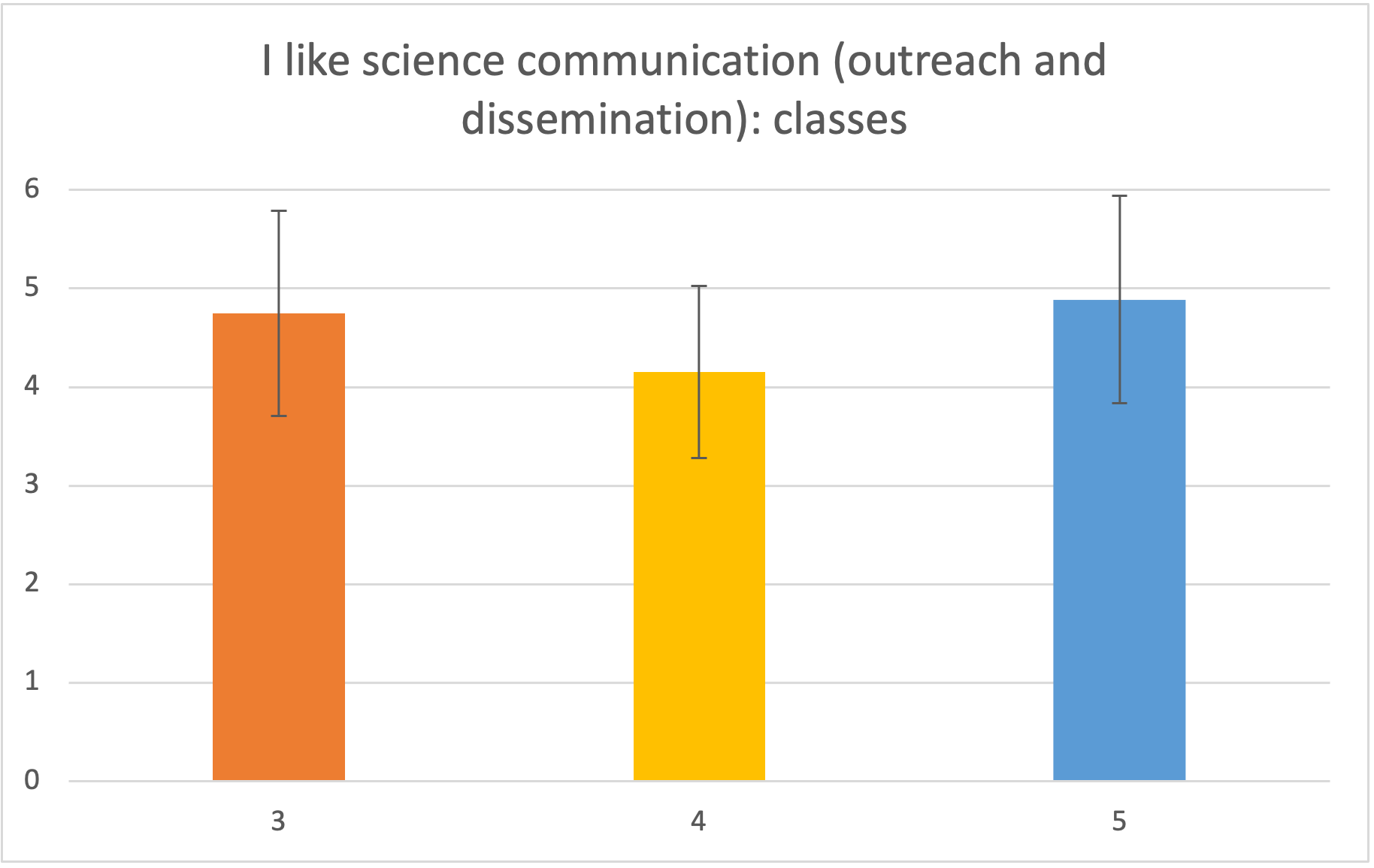}
    %\caption{The panel shows students' distribution as a function of the gender.}
  \end{minipage}
  \caption{The panel shows means and standard deviation (error bars) about students' interest in science communication according to the type of schools (left) and classes (right). Students rated using a 6-point Likert scale, from 1 (completely disagree) to 6 (completely agree).}\label{fig:outreach_school_classes}
\end{figure}

\begin{figure}[h]
\centering%
\includegraphics[width=0.5\textwidth]{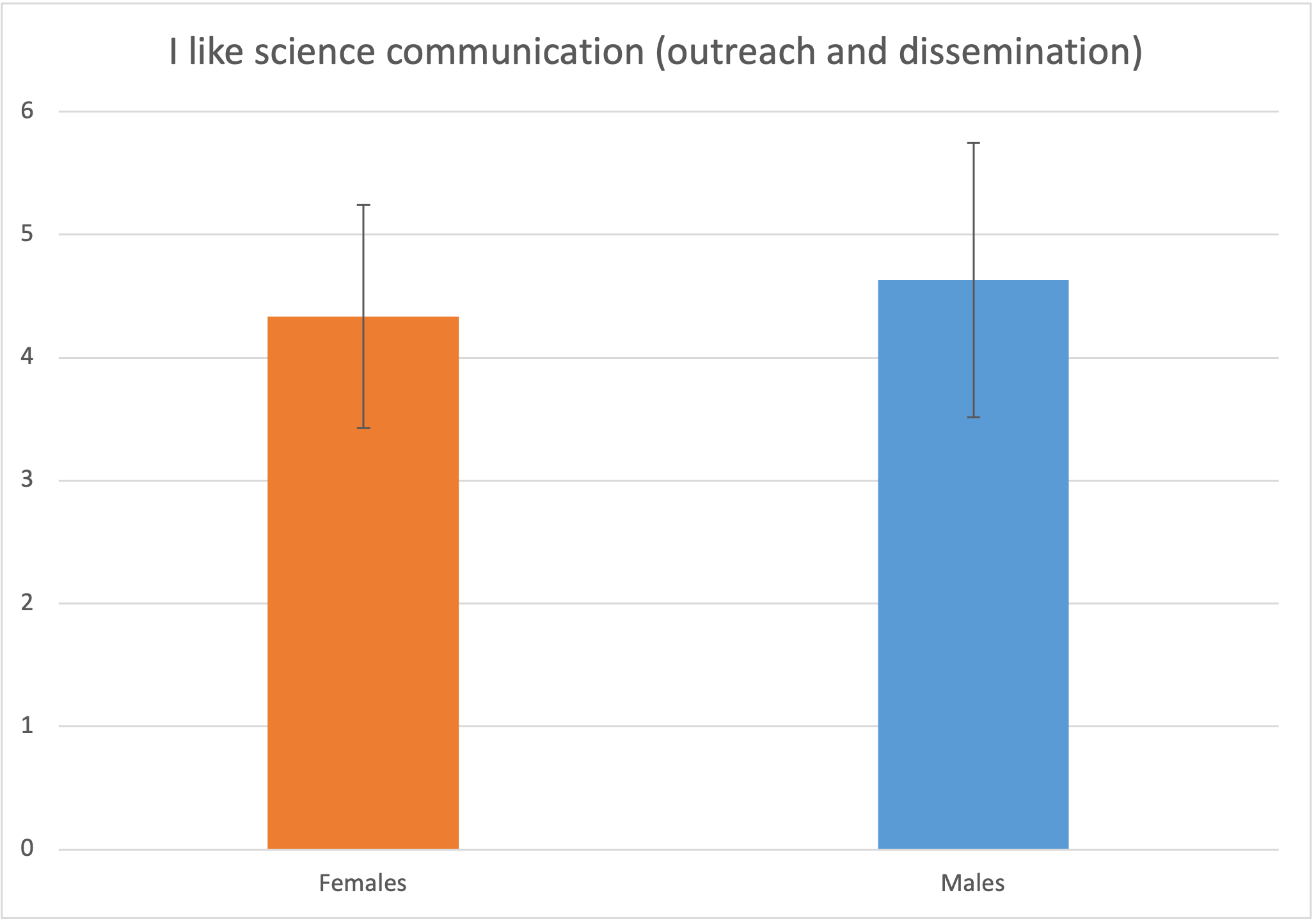}  
\caption{The panel shows means and standard deviation (error bars) about students' interest in science communication according to the gender. Students rated using a 6-point Likert scale, from 1 (completely disagree) to 6 (completely agree).}\label{fig:school_outreach}
\end{figure}

Concerning STEM Engagement and interest (from now-on \lq\lq STEM Engagement\rq\rq), we asked if students are interested in science communication, i.e. in outreach and dissemination activities: 1 student (1.3\%) is not interested at all, 12 (15.8\%) are indifferent to such activities, 27 (35.5\%) are slightly interested, 23 (30.3\%) are interested and 13 (17.1\%) are very interested in them, see Fig.~\ref{fig:interest_outreach}. Concerning the schools and the classes means and standard deviation are shown in Fig.~\ref{fig:outreach_school_classes}. Finally, we also reported gender results in Fig.~\ref{fig:school_outreach}.

Students also gave their feedback about the role of science communication and outreach in society. In particular, we asked them if we need more science communication and outreach activities at school: 65 (85.5\%) students answered \lq\lq yes\rq\rq, 1 (1.3\%) said \lq\lq no\rq\rq~and 10 (13.2\%) \lq\lq I do not know\rq\rq. We also asked them if we need more science communication and outreach activities locally, all around the Region: 53 (69.7\%) students rated \lq\lq yes\rq\rq, 3 (3.9\%) answered \lq\lq no\rq\rq~and 10 (26.3\%) \lq\lq I do not know\rq\rq. Finally, we collected their feedback about the need for more science communication and outreach activities in our society: 65 (85.5\%) answered \lq\lq yes\rq\rq, 2 (2.6\%) \lq\lq no\rq\rq, and 9 (11.8\%) \lq\lq I do not know\rq\rq. 

%OUTREACH TOOLS
%%
\begin{figure}[h]
\centering%
\includegraphics[width=0.6\textwidth]{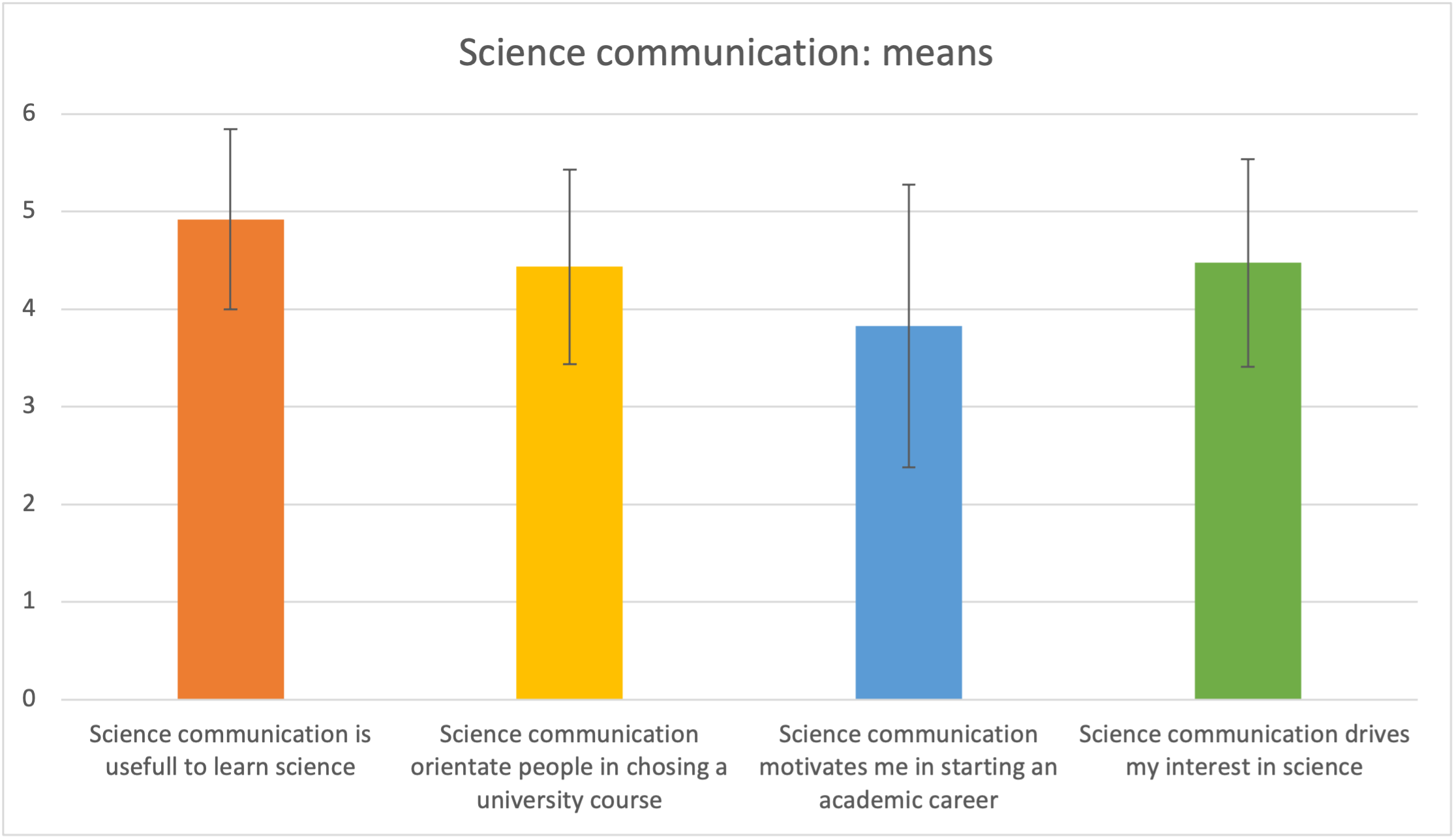}  
\caption{The panel shows means and standard deviation (error bars) about the use of science communication as a tool to engage students in STEM.}\label{fig:stem}
\end{figure}

\begin{figure}[h]
\centering%
\includegraphics[width=0.6\textwidth]{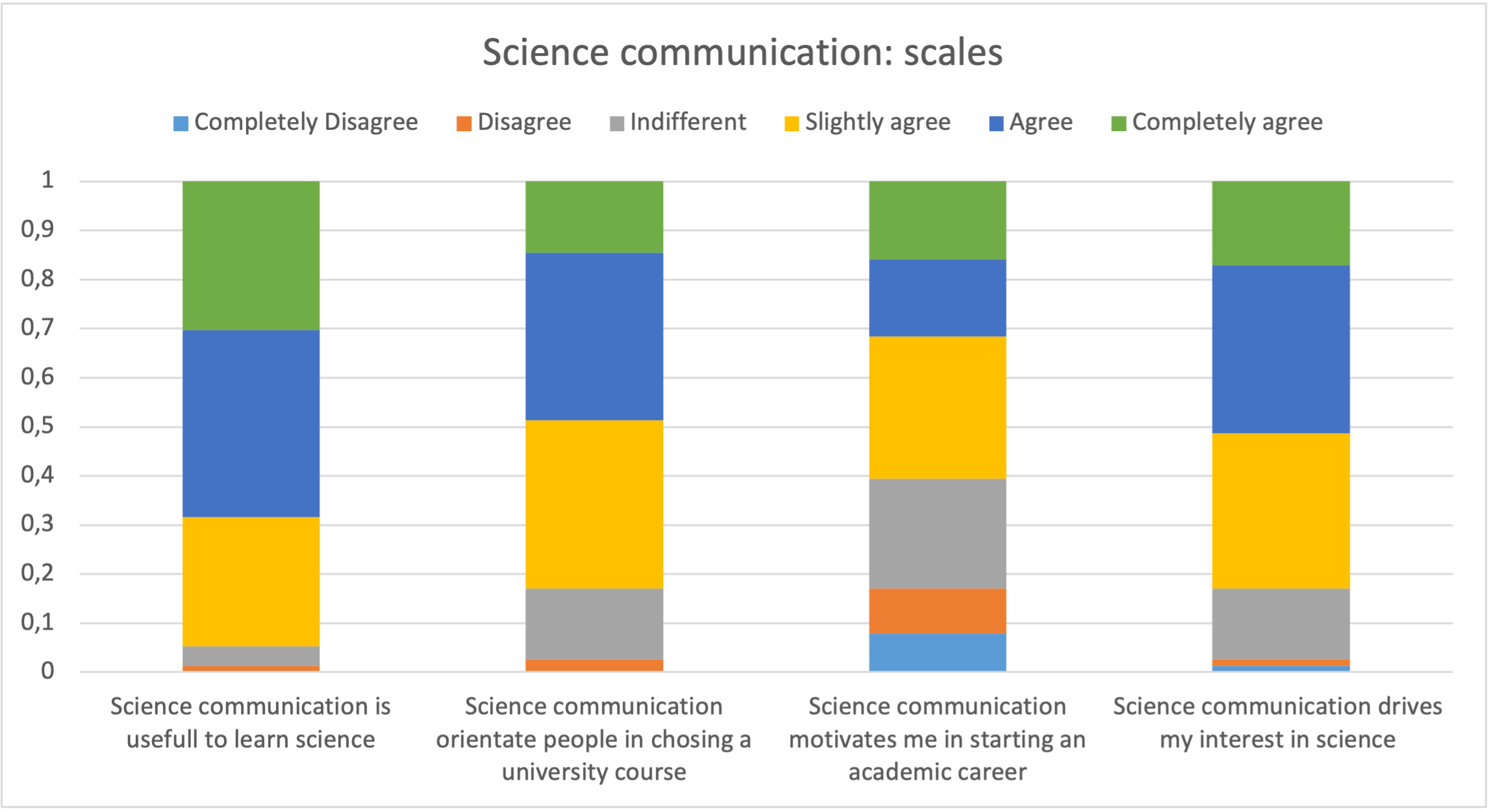}  
\caption{The panel shows students' feedback on the use of science communication as a tool to engage students in STEM according to the 6-point Likert scale.}\label{fig:stem_scales}
\end{figure}

\begin{figure}[htbp]
  \centering
  \begin{minipage}[b]{0.6\textwidth}
    \includegraphics[width=\textwidth]{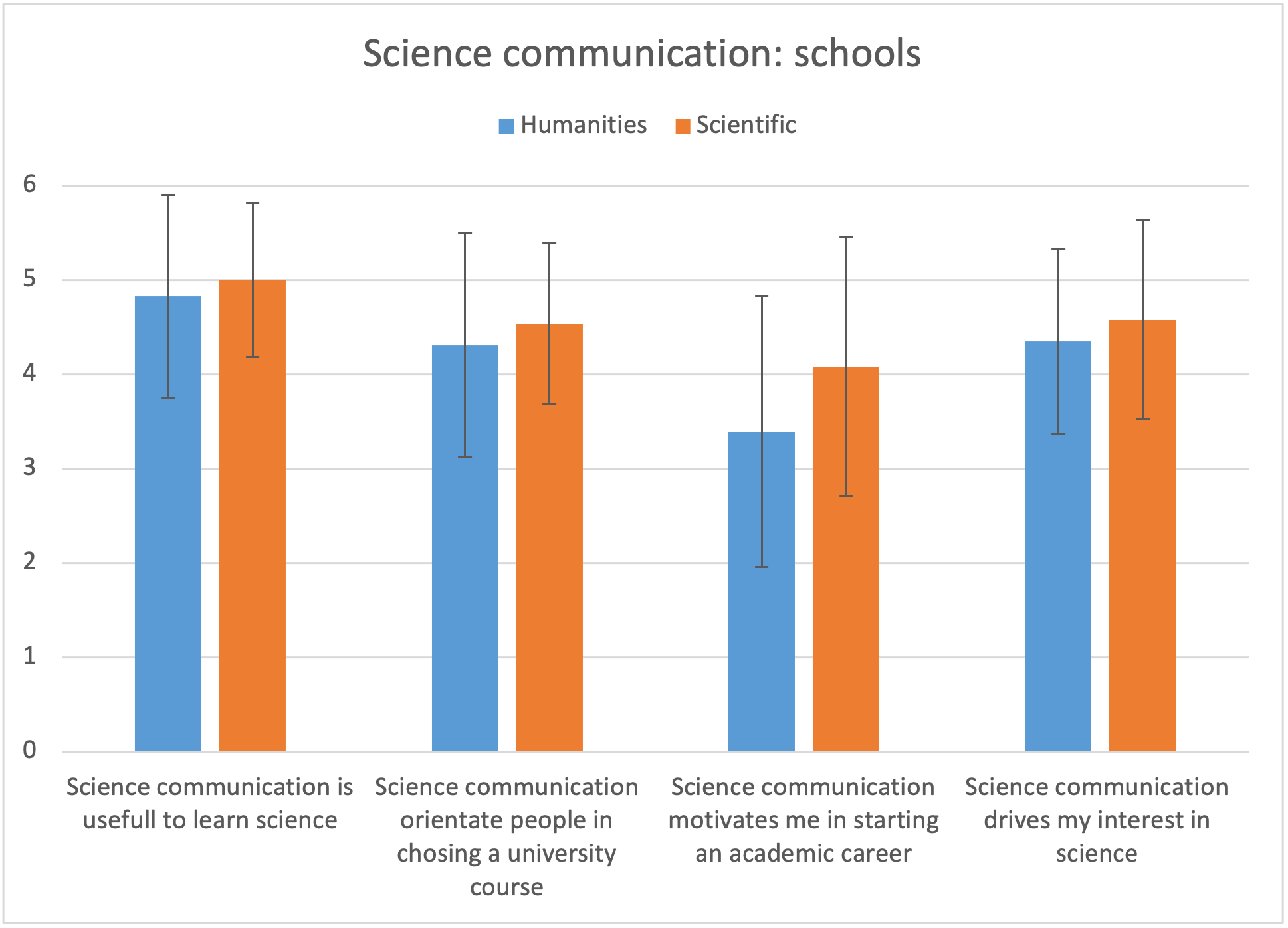}
    %\caption{The panel shows students' distribution as a function of the classes.}
  \end{minipage}
  \hfill
  \begin{minipage}[b]{0.6\textwidth}
    \includegraphics[width=\textwidth]{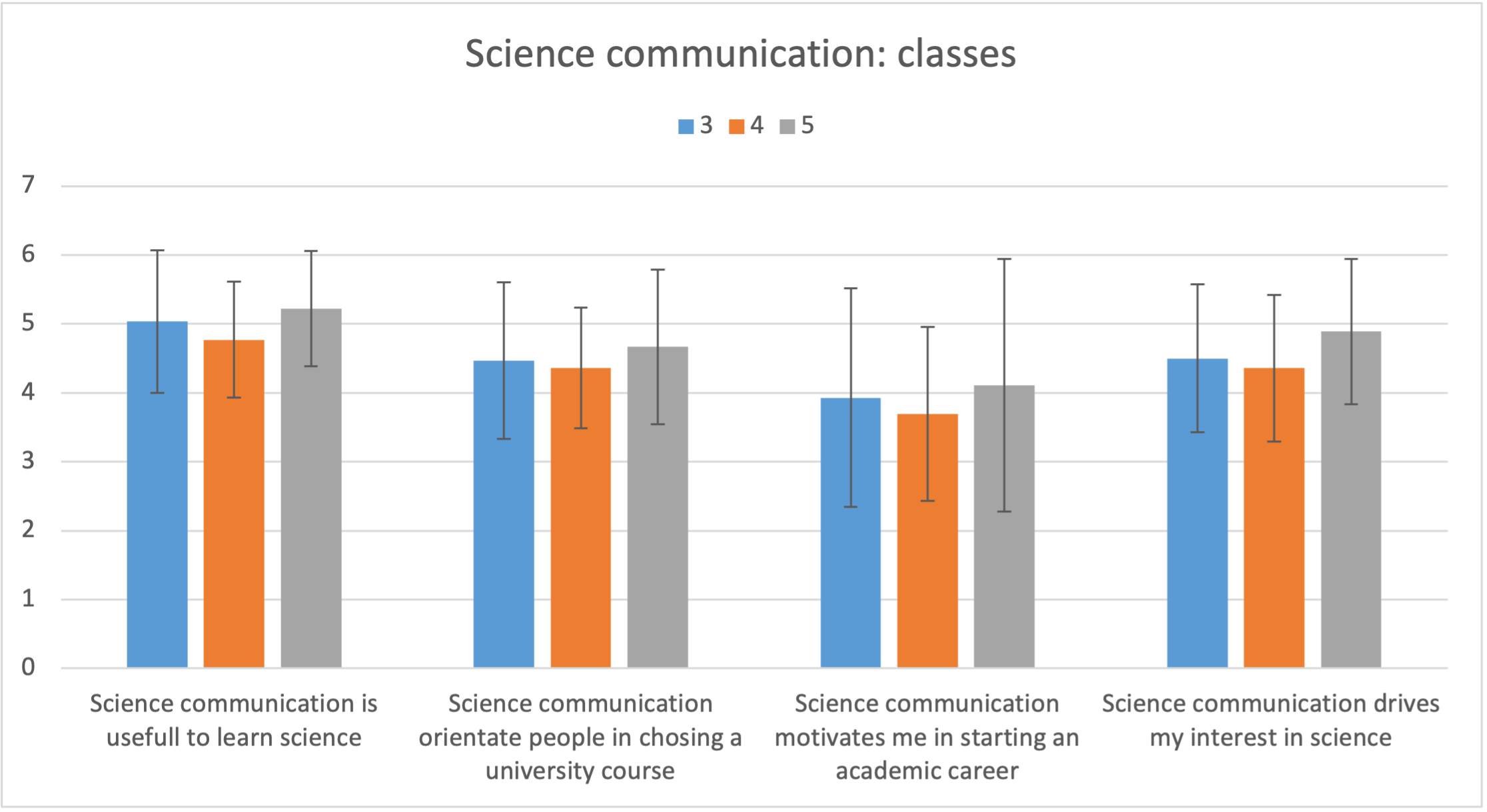}
    %\caption{The panel shows students' distribution as a function of the gender.}
  \end{minipage}
  \caption{The panel shows means and standard deviation (error bars) about the use of science communication as a tool to engage students in STEM according to the type of school (above) and the classes (below). }\label{fig:stem_school_classes}
\end{figure}
\begin{figure}[h]
\centering%
\includegraphics[width=0.6\textwidth]{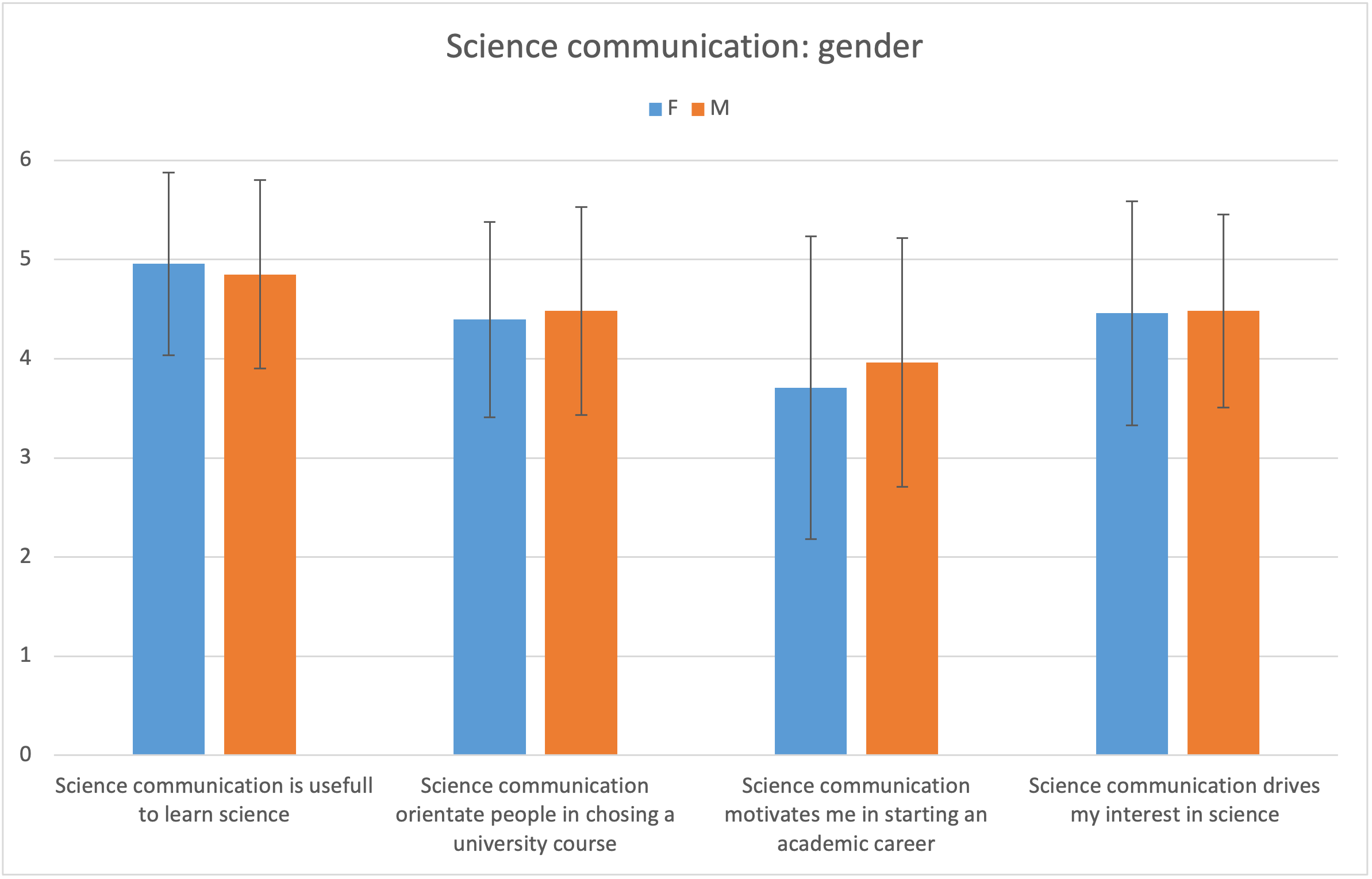}  
\caption{The panel shows means and standard deviation (error bars) about students' interest in science communication according to the gender.}\label{fig:stem_gender}
\end{figure}

We also investigated students' feedback about the role of science communication (outreach and dissemination) as a tool to learn science (a), to orientate people in choosing (b) and pursuing (c) an academic career. Finally, we also asked them if science communication drives their interest in science. Students rated using a 6-point Likert scale, from 1 (completely disagree) to 6 (completely agree). Results are shown in Fig.~\ref{fig:stem_scales}. 
Results concerning means and standard deviation for the four items are shown in Fig.~\ref{fig:stem}. In Figs~\ref{fig:stem_school_classes} and~\ref{fig:stem_gender}, we show results divided per school, classes, and gender, respectively.

%%THE MONOLOGUE 
%%
\begin{figure}[h]
\centering%
\includegraphics[width=0.8\textwidth]{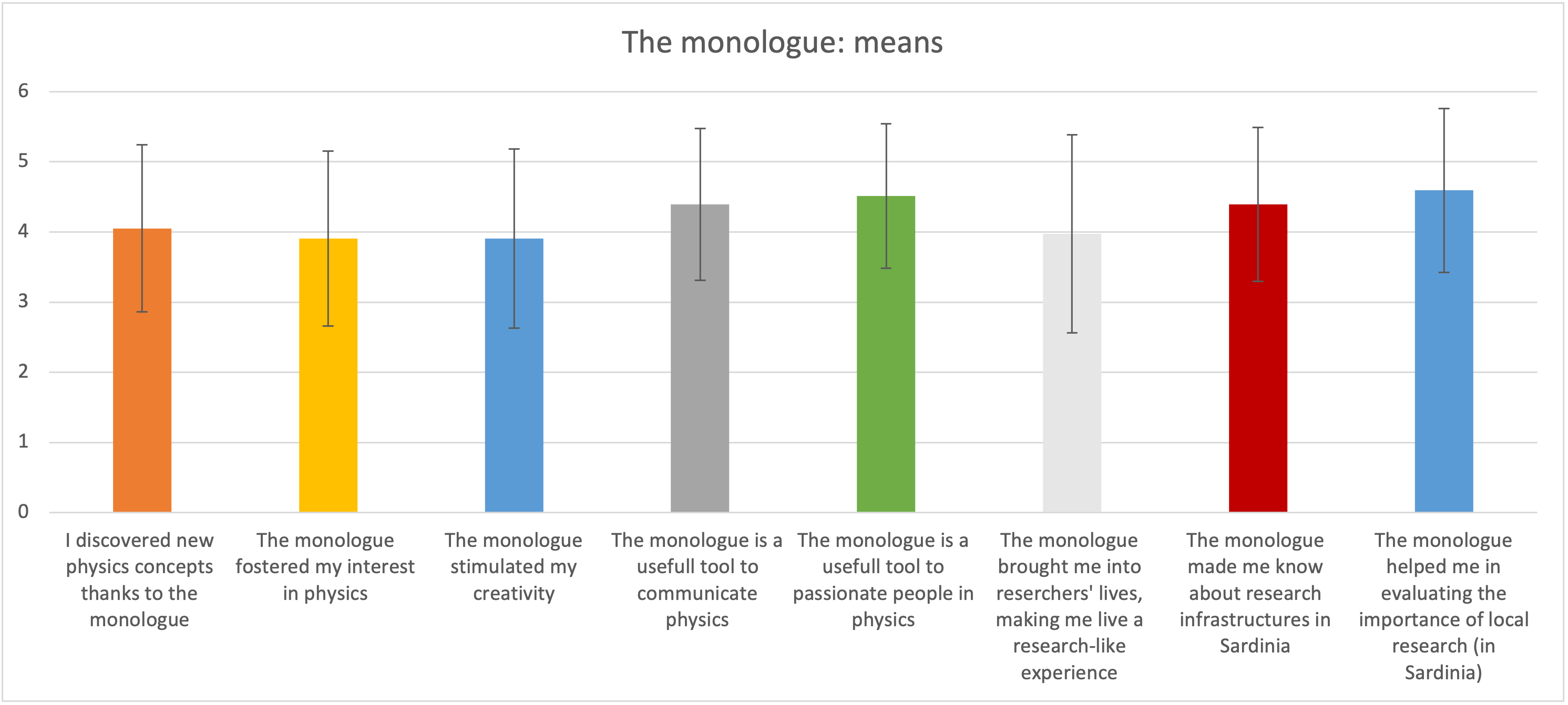}  
\caption{The panel shows means and standard deviation (error bars) about the use of the monologue to engage students' in physics. Students rated using a 6-point Likert scale, from 1 (completely disagree) to 6 (completely agree).}\label{fig:monologue}
\end{figure}

\begin{figure}[htbp]
  \centering
  \begin{minipage}[b]{0.9\textwidth}
    \includegraphics[width=\textwidth]{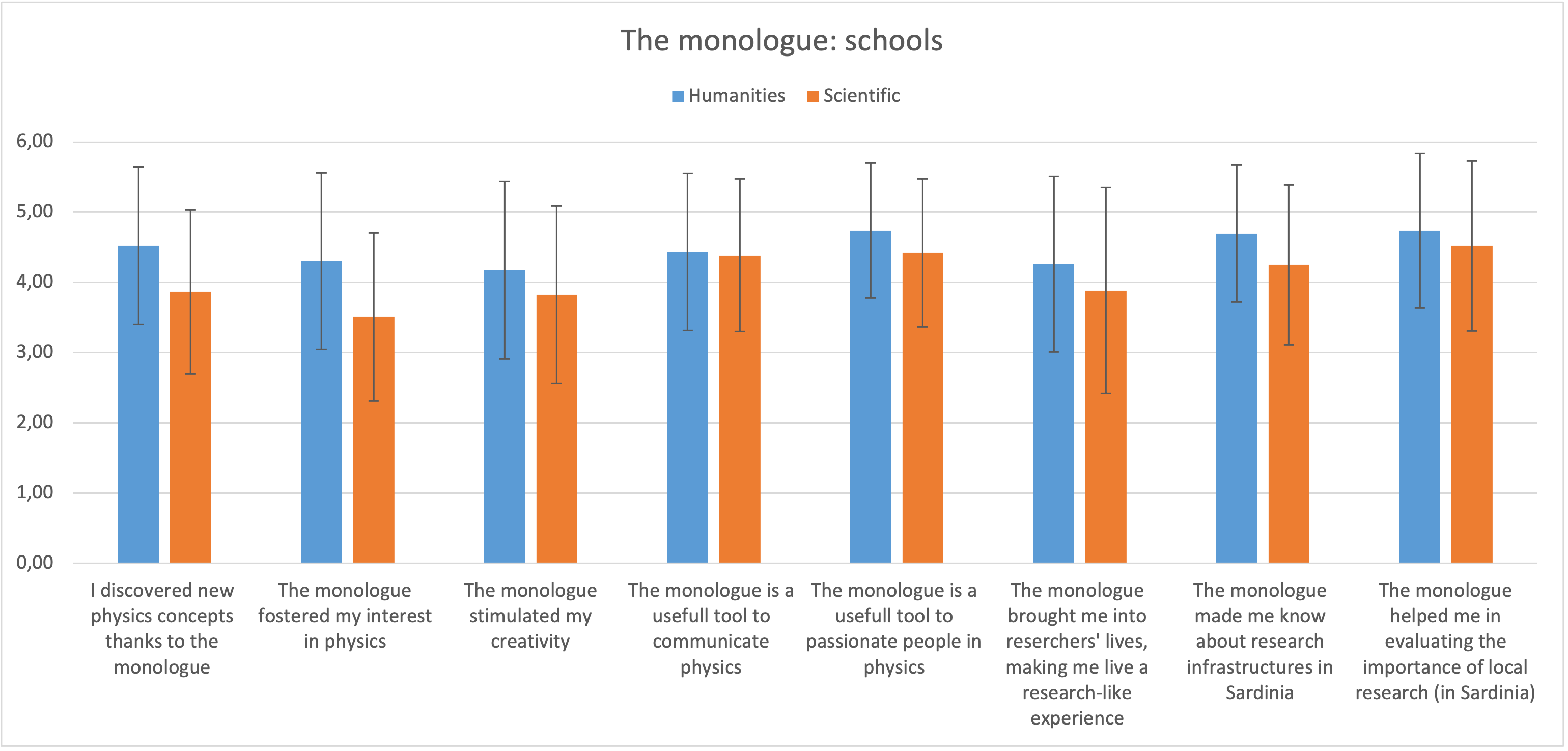}
    %\caption{The panel shows students' distribution as a function of the classes.}
  \end{minipage}
  \hfill
  \begin{minipage}[b]{0.9\textwidth}
    \includegraphics[width=\textwidth]{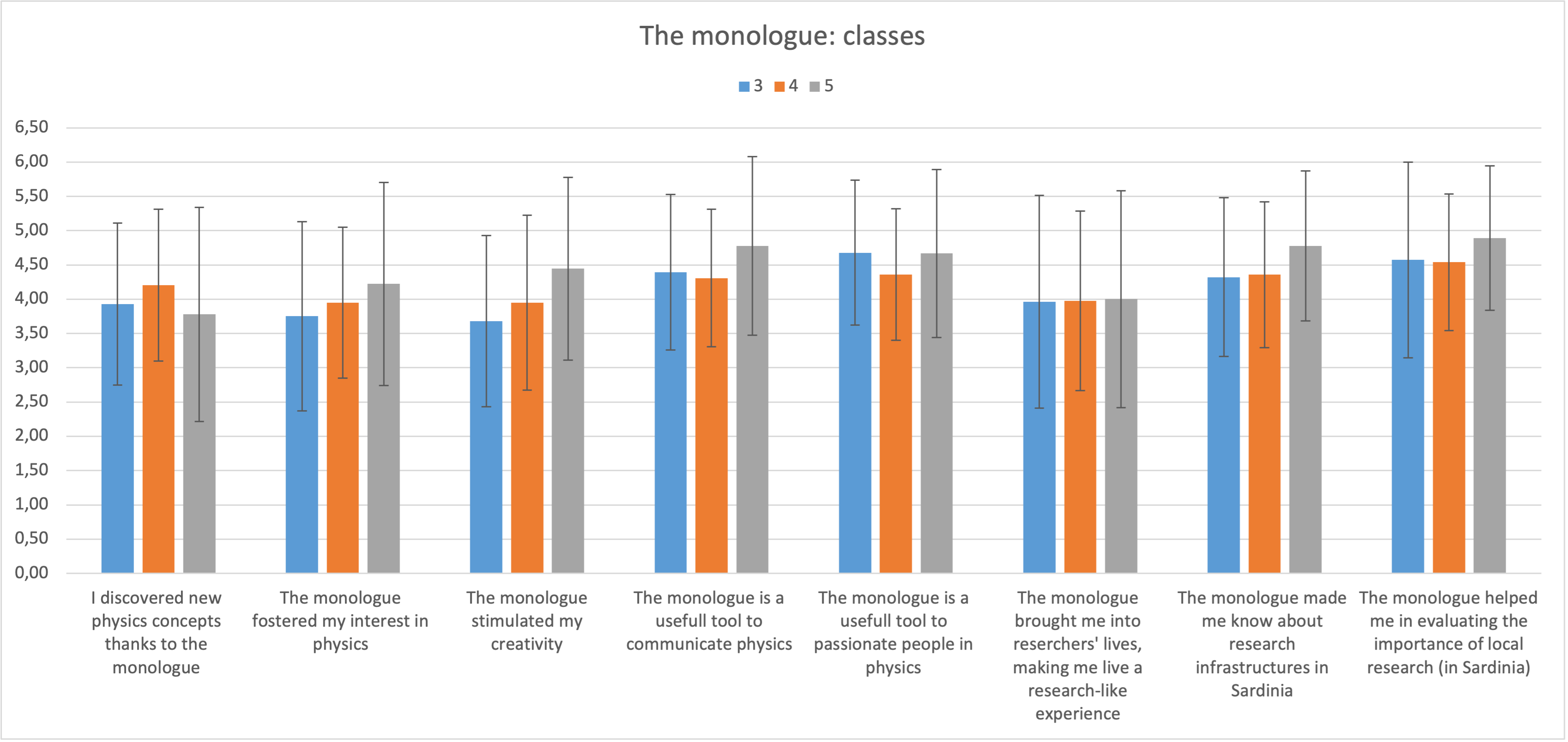}
    %\caption{The panel shows students' distribution as a function of the gender.}
  \end{minipage}
  \caption{The panel shows means and standard deviation (error bars) about the use of science communication as a tool to engage students in STEM according to the type of school (above) and the classes (below). }\label{fig:monologue_school_classes}
\end{figure}
\begin{figure}[h]
\centering%
\includegraphics[width=1.0\textwidth]{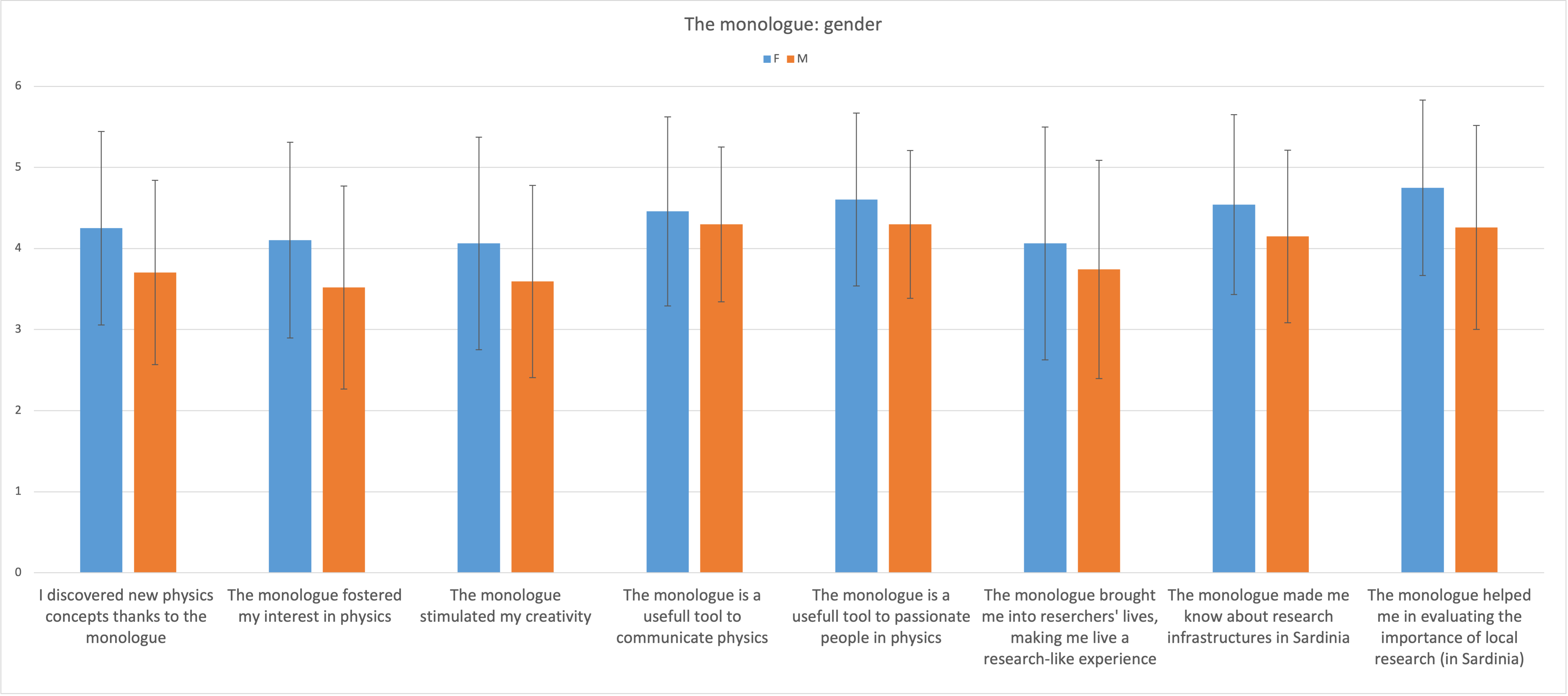}  
\caption{The panel shows means and standard deviation (error bars) about students' interest in science communication according to the gender. Students rated using a 6-point Likert scale, from 1 (completely disagree) to 6 (completely agree).}\label{fig:monologue_gender}
\end{figure}

Concerning the macro-domain related to the monologue, we investigated 8 different items to study the efficacy of the artistic tool on learning, motivation, engagement, and the presence of research infrastructure in Sardinia. The items were the following: a) I discovered new physics concepts thanks to the monologue; b)  The monologue fostered my interest in physics; c) The monologue stimulated my creativity; d) The monologue is a useful tool to communicate physics; e) The monologue is a useful tool to passionate people in physics; f) The monologue brought me into researchers' lives, making me live a research-like experience; g) The monologue made me know about research infrastructures in Sardinia: h) The monologue helped me in evaluating the importance of local research (in Sardinia). Also in this case, students could answer using a 6-point Likert scale, from 1 (completely disagree) to 6 (completely agree), see Table~\ref{table:monologue}. 
Means and standard deviation, school, and class results are shown in Figs.~\ref{fig:monologue} and~\ref{fig:monologue_school_classes}. Gender results are shown in Fig.~\ref{fig:monologue_gender}.

\begin{table}[h]\label{table:monologue}
\begin{tiny}
\begin{center}
\begin{tabular}{ | p{2.4cm}| | p{2cm}| p{1.8cm}| p{1.5cm}| p{1.8cm}| p{1.5cm}| p{1.8cm}| }
 \hline
 \multicolumn{7}{|c|}{The Monologue} \\
 \hline
  &  Completely Disagree & Disagree & Indifferent & Slightly Agree & Agree & Completely agree \\
 \hline
 I discovered new physics concepts thanks to the monologue  & 1 (1.3\%) & 8 (10.5\%)  & 11 (14.5\%)  & 32 (42.1\%) & 14 (18.0\%) & 10 (13.2\%)  \\
 & & & & & & \\
The monologue fostered my interest in physics & 1 (1.3\%)  & 12 (15.8\%)  &  12 (15.8\%) & 27 (35.5\%) & 16 (21.0\%) & 8 (10.5\%) \\
& & & & & & \\
 The monologue stimulated my creativity & 1 (1.0\%) & 11 (14.5\%) & 18 (23.7\%) & 18 (23.7\%) & 20 (26.7\%) &8 (10.5\%) \\
 & & & & & & \\
 The monologue is a usefull tool to communicate physics & 0 (0\%) & 5 (6.6\%) & 7 (9.2\%) & 30 (39.5\%) &  21 (27.6\%) & 13 (17.1\%)  \\
 & & & & & & \\
 The monologue is a usefull tool to passionate people in physics & 0 (0\%) & 2 (2.6\%) & 8 (10.5\%) & 31 (40.8\%) & 16 (25.0\%) & 19 (25.3\%) \\
 & & & & & & \\
 The monologue brought me into researchers' lives, making me live a research-like experience & 4 (5.3\%) & 10 (13.2\%) & 11 (14.5\%) & 21 (27.6\%) & 19 (25.0\%) & 11 (14.5\%)  \\
 & & & & & & \\
 The monologue made me know about research infrastructures in Sardinia & 2 (2.6\%) & 2 (2.6\%) & 7 (9.2\%) & 29 (38.2\%)  & 25 (32.9\%)  & 11 (14.5\%)  \\
 & & & & & & \\
 The monologue helped me in evaluating the importance of local research (in Sardinia) & 2 (2.6\%) & 2 (2.6\%) & 7 (9.2\%) & 20 (26.3\%) & 28 (36.8\%)  & 17 (22.3\%) \\  
  \hline
  \end{tabular}
\end{center}
\caption{The table shows the number of answers (and corresponding percentage) for each point of the Likert scale concerning students' feedback about the efficacy of the monologue on the selected domains.}
\end{tiny}
\end{table}

%%POEMS

\begin{figure}[h]
\centering%
\includegraphics[width=0.7\textwidth]{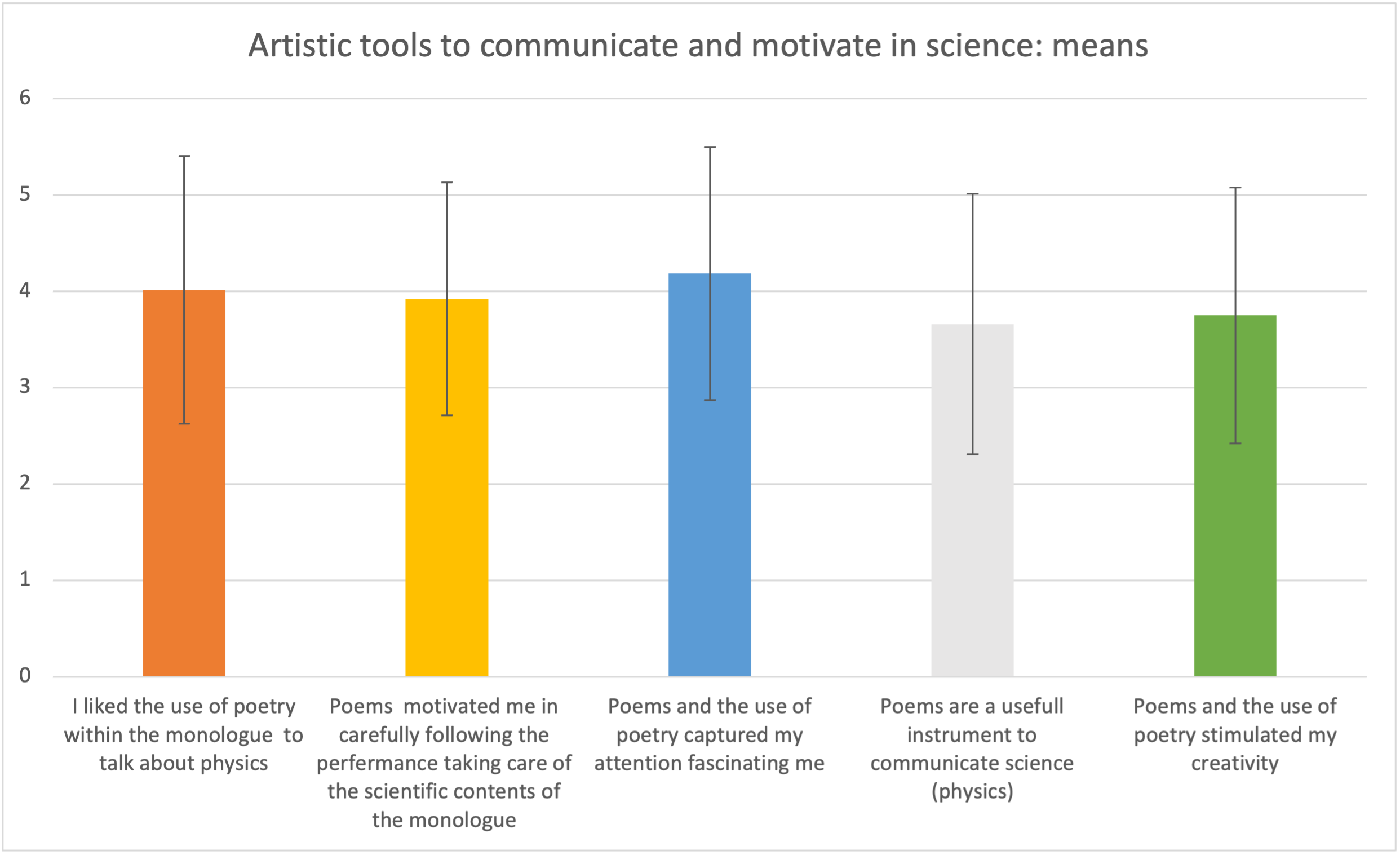}  
\caption{The panel shows means and standard deviation (error bars) about the use of poems within the monologue as an educational tool.}\label{fig:poems}
\end{figure}

\begin{figure}[htbp]
  \centering
  \begin{minipage}[b]{0.7\textwidth}
    \includegraphics[width=\textwidth]{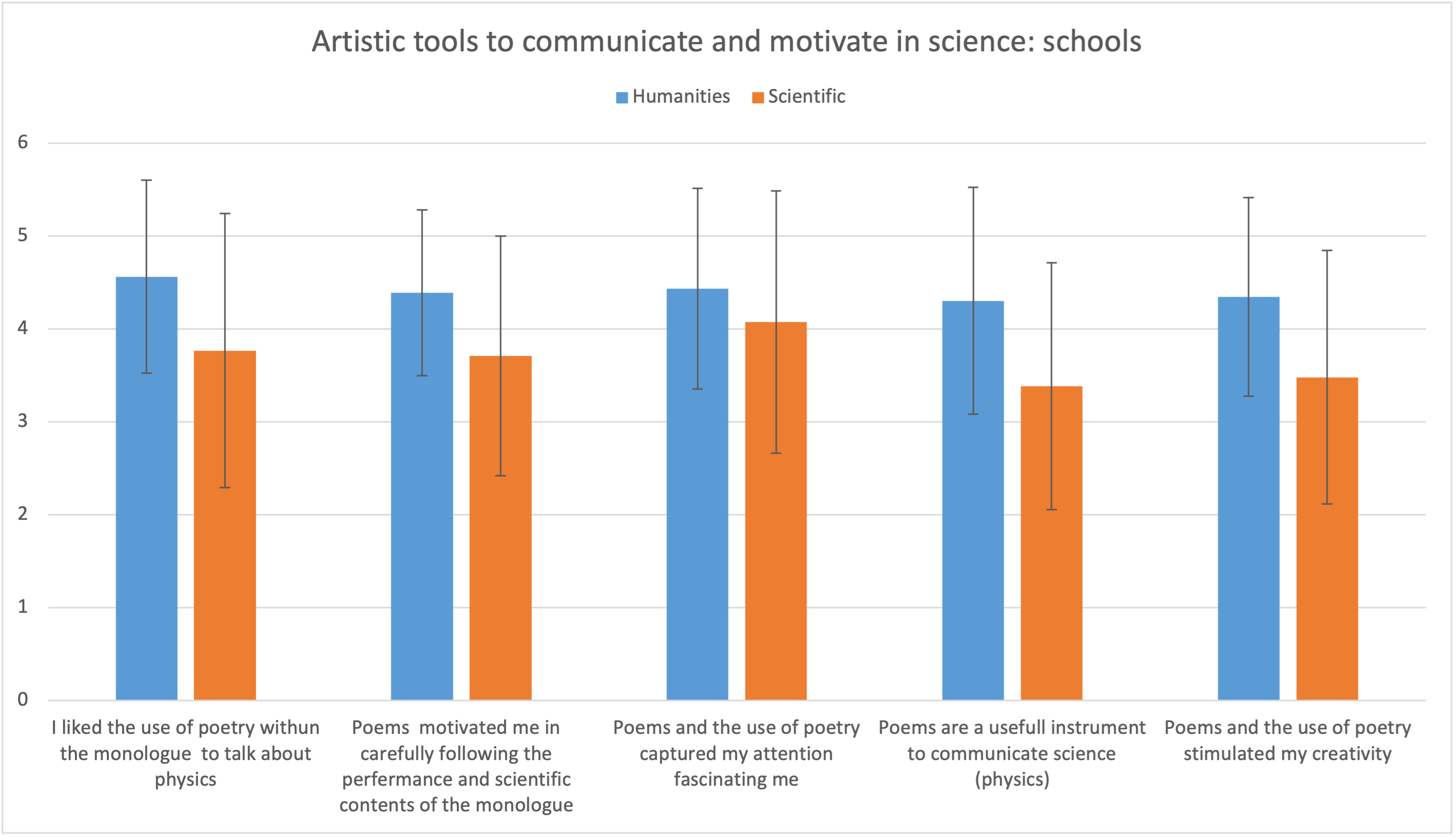}
    %\caption{The panel shows students' distribution as a function of the classes.}
  \end{minipage}
  \hfill
  \begin{minipage}[b]{0.7\textwidth}
    \includegraphics[width=\textwidth]{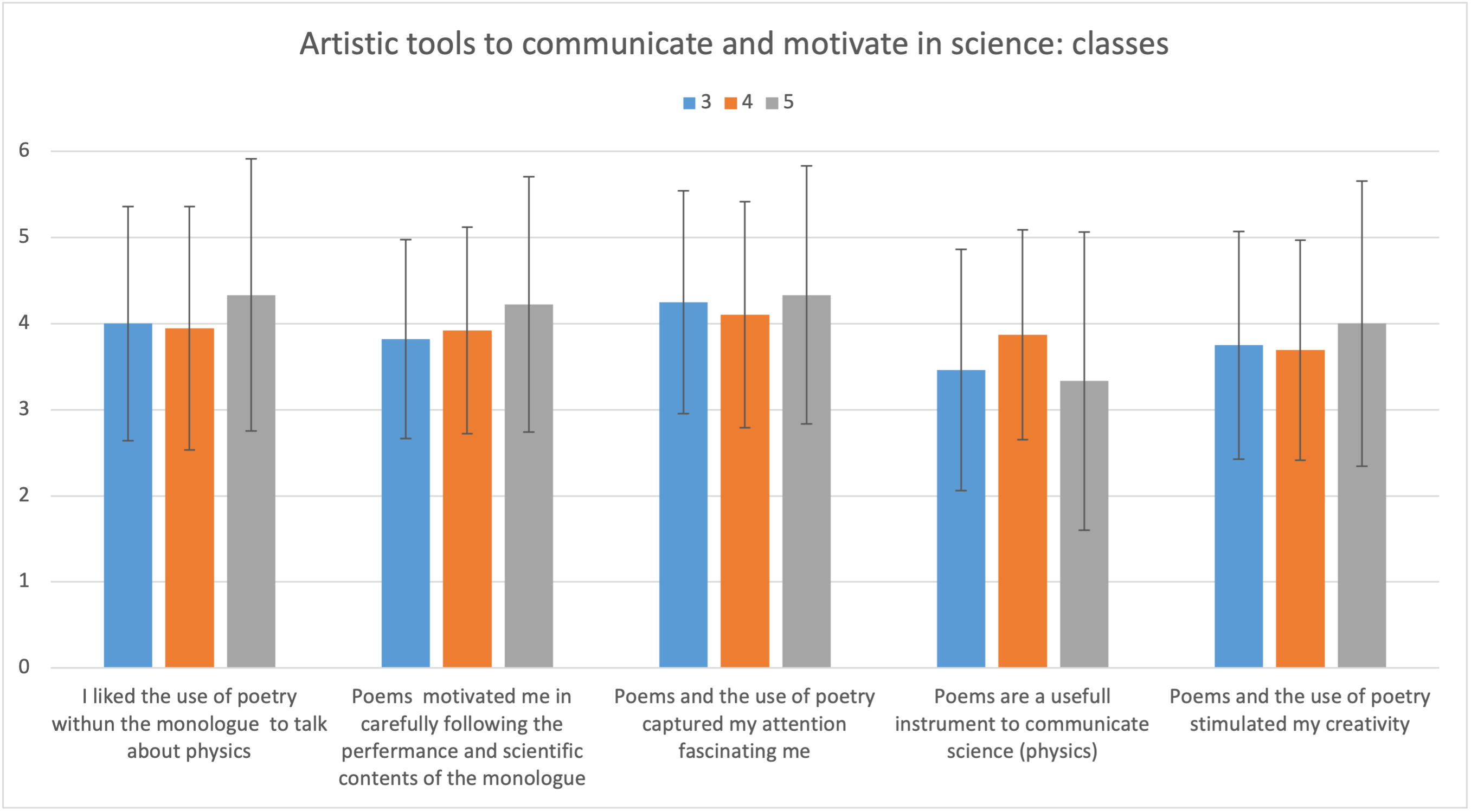}
    %\caption{The panel shows students' distribution as a function of the gender.}
  \end{minipage}
  \caption{The panel shows means and standard deviation (error bars) about the use of poems within the monologue as an educational tool according to the type of school (above) and the classes (below). }\label{fig:poems_school_classes}
\end{figure}
\begin{figure}[h]
\centering%
\includegraphics[width=0.7\textwidth]{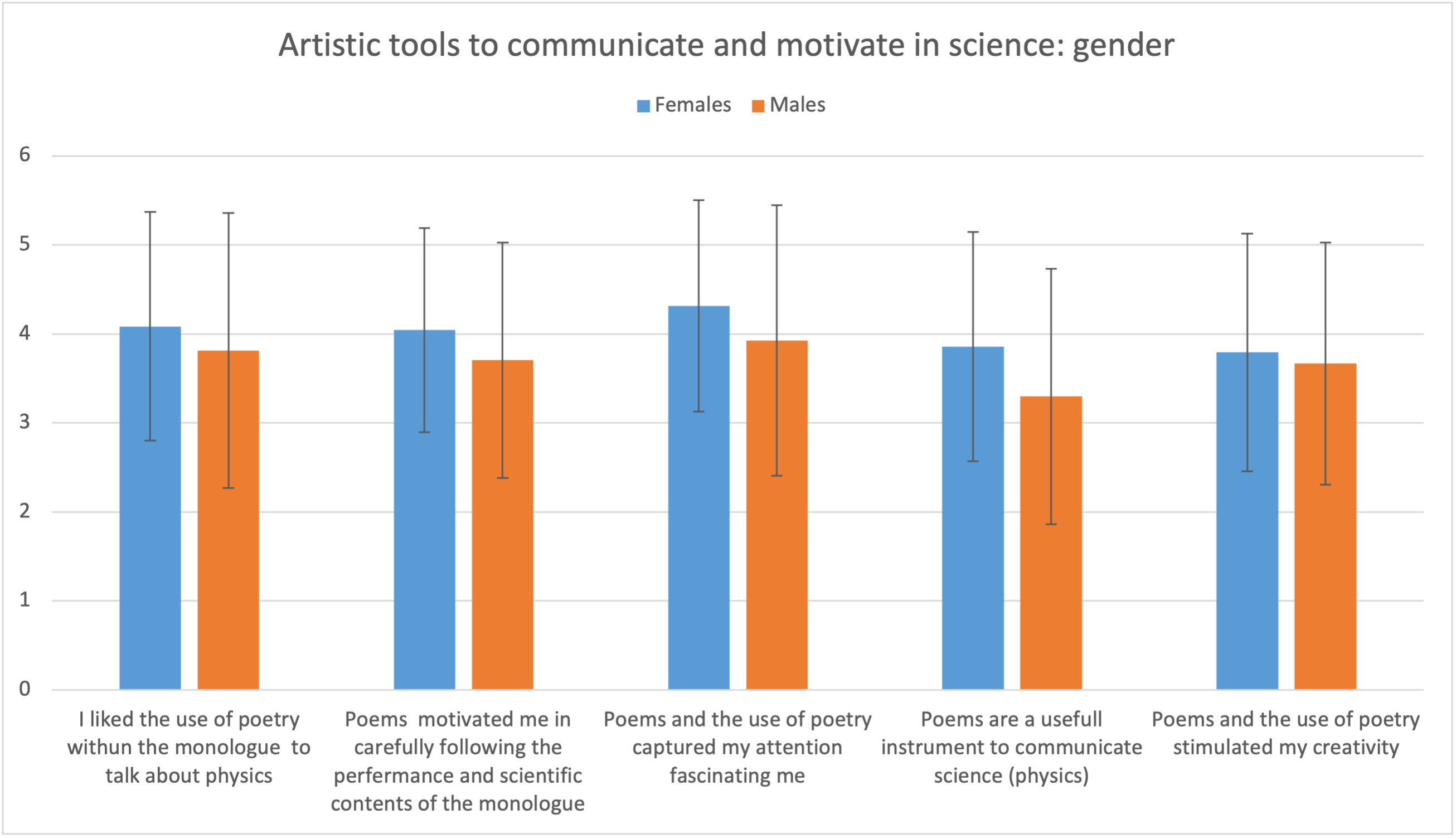}  
\caption{The panel shows means and standard deviation (error bars) about the use of poems within the monologue as an educational tool according to the gender.}\label{fig:poems_gender}
\end{figure}

Concerning the use of poems as an educational tool, we asked students if a) they liked the use of poetry within the monologue to talk about physics; b) poems motivated them to carefully follow the performance, taking care of the scientific contents of the monologue; c) poems and the use of poetry captured their attention, fascinating them; d) poems are a useful instrument to communicate science (physics); e) poems and the use of poetry stimulated their creativity. Also in this case students rated items using a 6-point Likert scale, from 1 (completely disagree) to 6 (completely agree). The means and standard deviation for each item are shown in Fig.~\ref{fig:poems}. We report results on single items according to the Likert scale in Table~\ref{table:poems}. School, classes, and gender results are shown in Figs.~\ref{fig:poems_school_classes} and~\ref{fig:poems_gender}.

\begin{table}[h]\label{table:poems}
\begin{tiny}
\begin{center}
\begin{tabular}{ | p{2.4cm}| | p{2cm}| p{1.8cm}| p{1.5cm}| p{1.8cm}| p{1.5cm}| p{1.8cm}| }
 \hline
 \multicolumn{7}{|c|}{Poems} \\
 \hline
  &  Completely Disagree & Disagree & Indifferent & Slightly Agree & Agree & Completely agree \\
 \hline
 I liked the use of poetry within the monologue to talk about physics  & 4 (5.2\%) & 10 (13.1\%)  & 7 (9.2\%)  & 26 (34.2\%) & 18 (23.7\%) & 11 (14.5\%)  \\
 & & & & & & \\
Poems motivated me in carefully following the performance, taking care of the scientific contents of the monologue & 3 (3.9\%)  & 6 (7.9\%)  &  16 (21.1\%) & 26 (34.2\%) & 19 (25.0\%) & 6 (7.9\%) \\
& & & & & & \\
 Poems and the use of poetry captured my attention, fascinating me & 4 (5.2\%) & 3 (3.9\%) & 13 (17.1\%) & 24 (31.6\%) & 19 (25.0\%) & 13 (17.1\%) \\
 & & & & & & \\
 Poems are a useful instrument to communicate science (physics) & 4 (5.2\%) & 11 (14.5\%) & 22 (28.9\%) & 16 (21.0\%) &  16 (21.0\%) & 7 (9.2\%)  \\
 & & & & & & \\
 Poems and the use of poetry stimulated my creativity & 6 (7.9\%) & 6 (7.9\%) & 19 (25.0\%) & 20 (26.3\%) & 20 (26.3\%) & 5 (6.6\%) \\ 
  \hline
  \end{tabular}
\end{center}
\caption{The table shows the number of answers (and corresponding percentage) for each point of the Likert scale concerning students' feedback about the use of poems as an educational tool.}
\end{tiny}
\end{table}

%MOTIVATION

\begin{figure}[h]
\centering%
\includegraphics[width=0.6\textwidth]{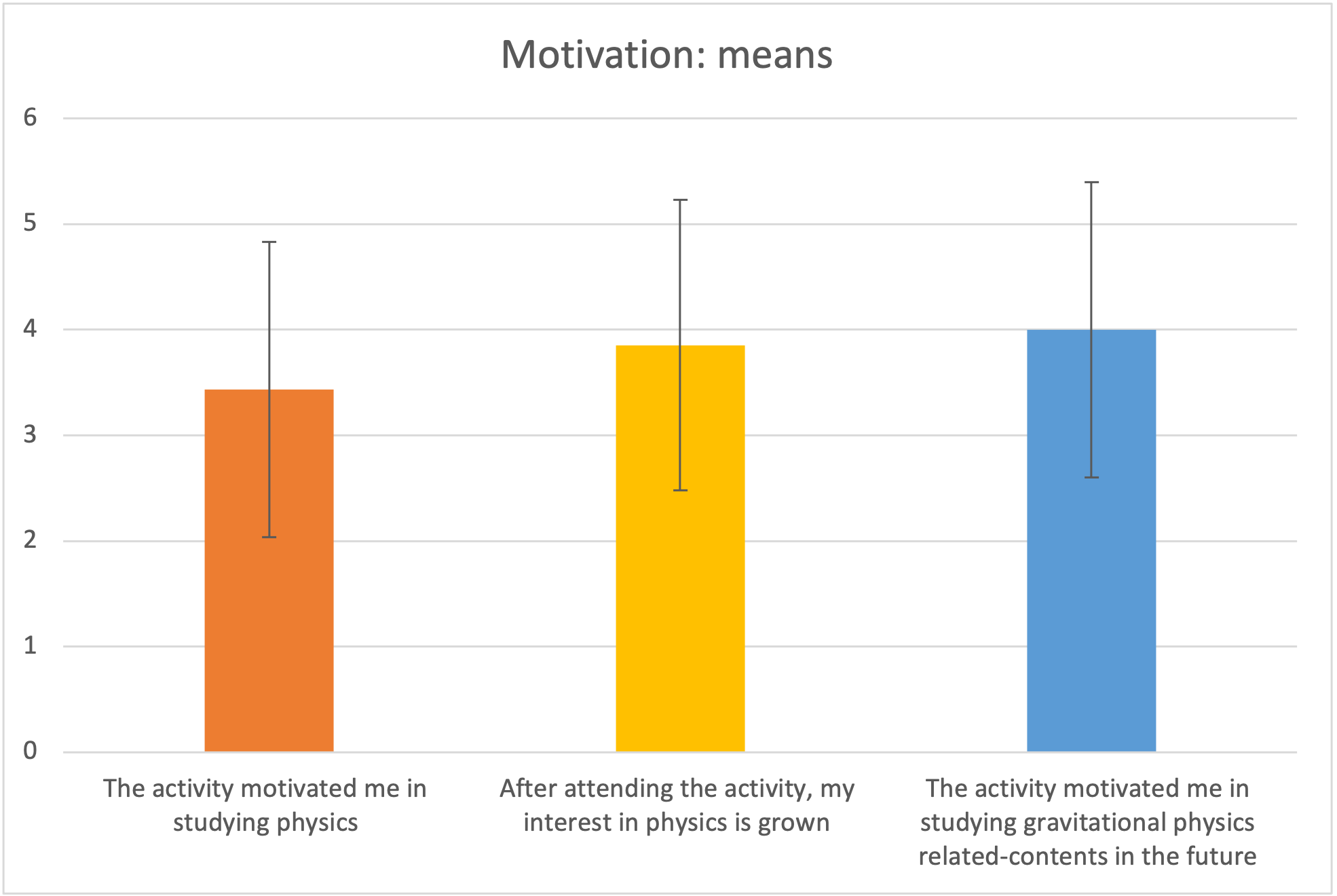}  
\caption{The panel shows means and standard deviation (error bars) about students' motivation in learning physics and the proposed subjects after attending the activity.}\label{fig:motivation}
\end{figure}

\begin{figure}[htbp]
  \centering
  \begin{minipage}[b]{0.6\textwidth}
    \includegraphics[width=\textwidth]{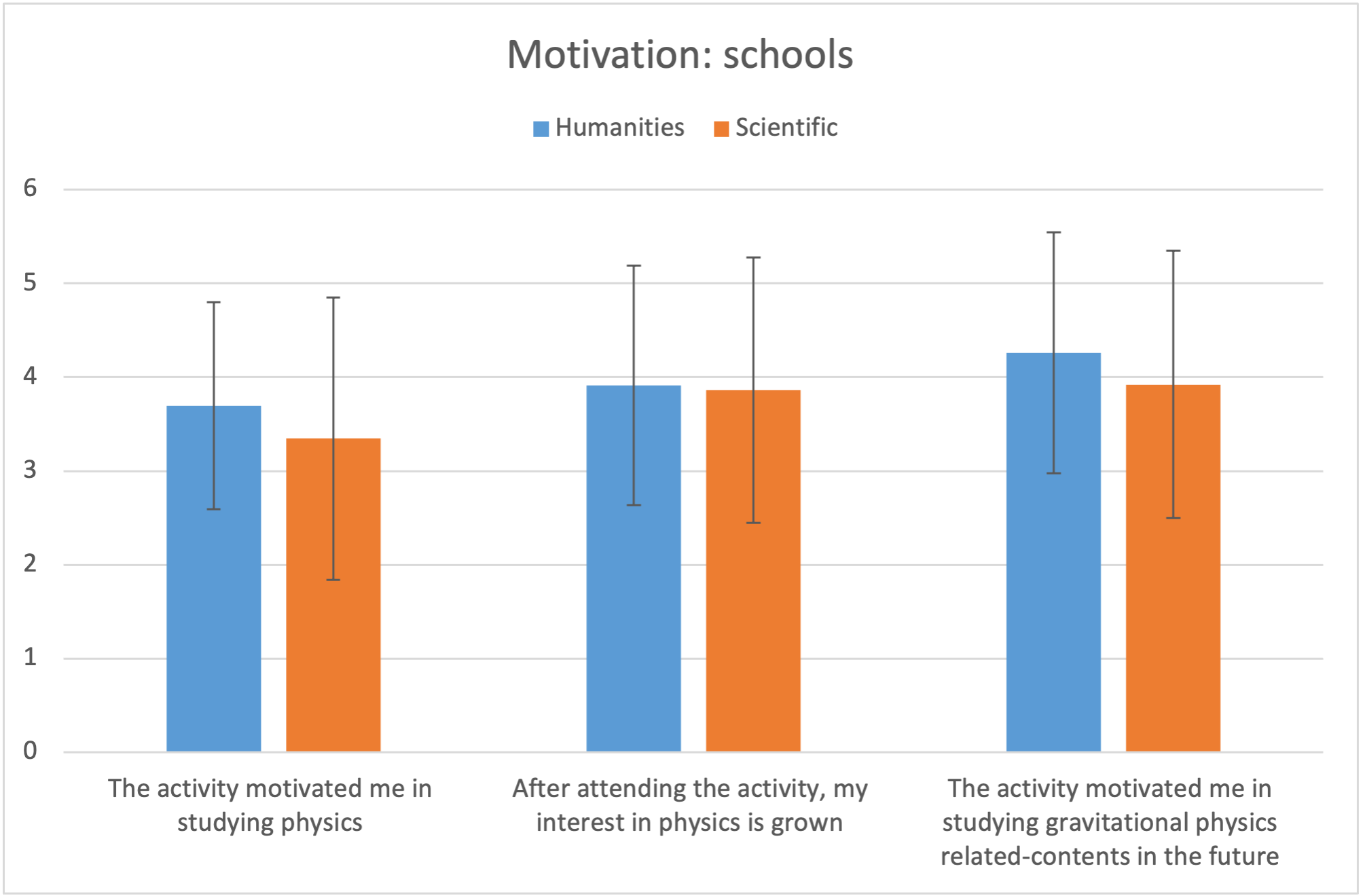}
    %\caption{The panel shows students' distribution as a function of the classes.}
  \end{minipage}
  \hfill
  \begin{minipage}[b]{0.6\textwidth}
    \includegraphics[width=\textwidth]{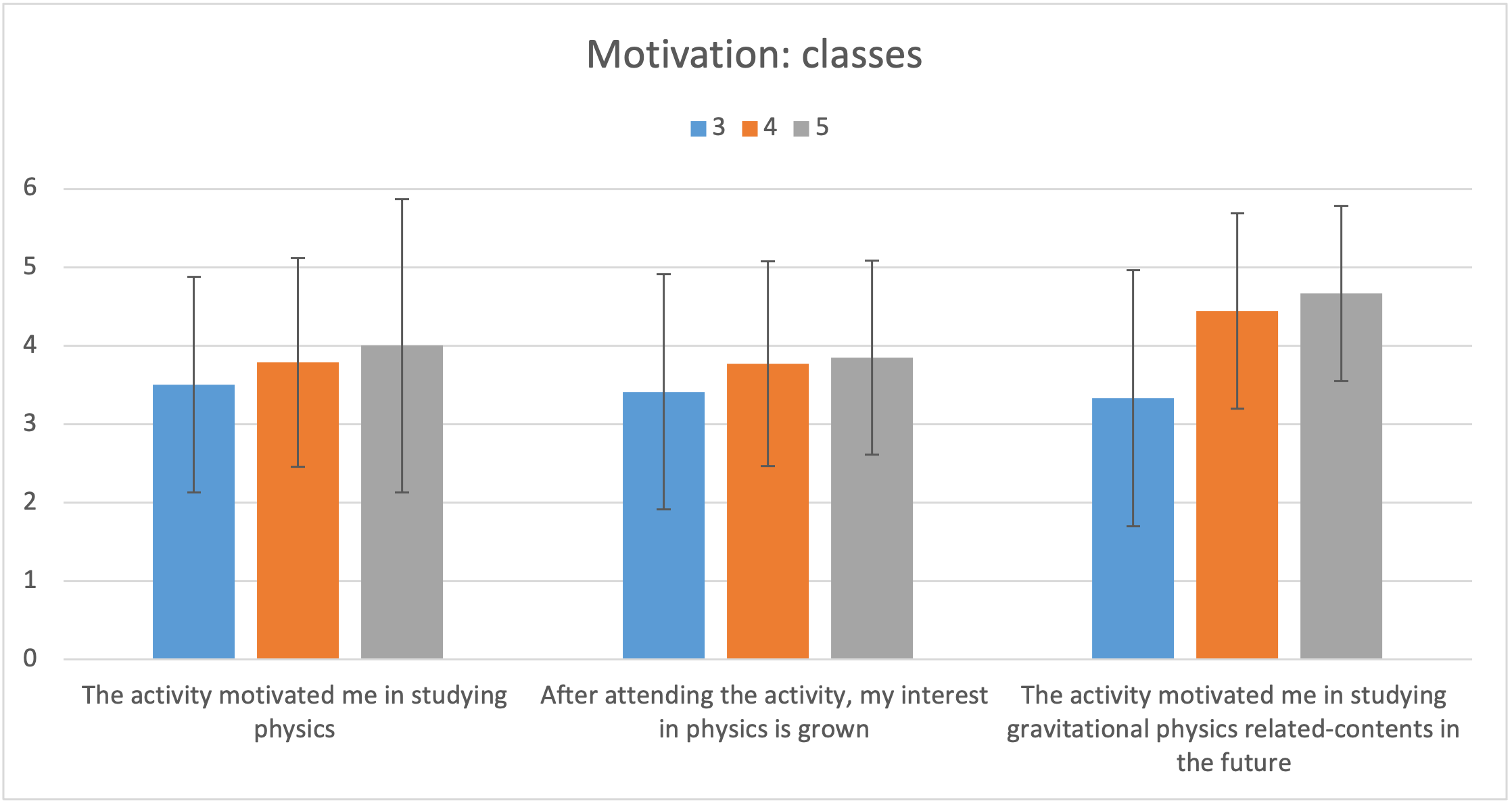}
    %\caption{The panel shows students' distribution as a function of the gender.}
  \end{minipage}
  \caption{The panel shows means and standard deviation (error bars) about students' motivation in learning physics and the proposed subjects after attending the activity according to the type of school (above) and the classes (below). }\label{fig:motivation_school_classes}
\end{figure}
\begin{figure}[h]
\centering%
\includegraphics[width=0.6\textwidth]{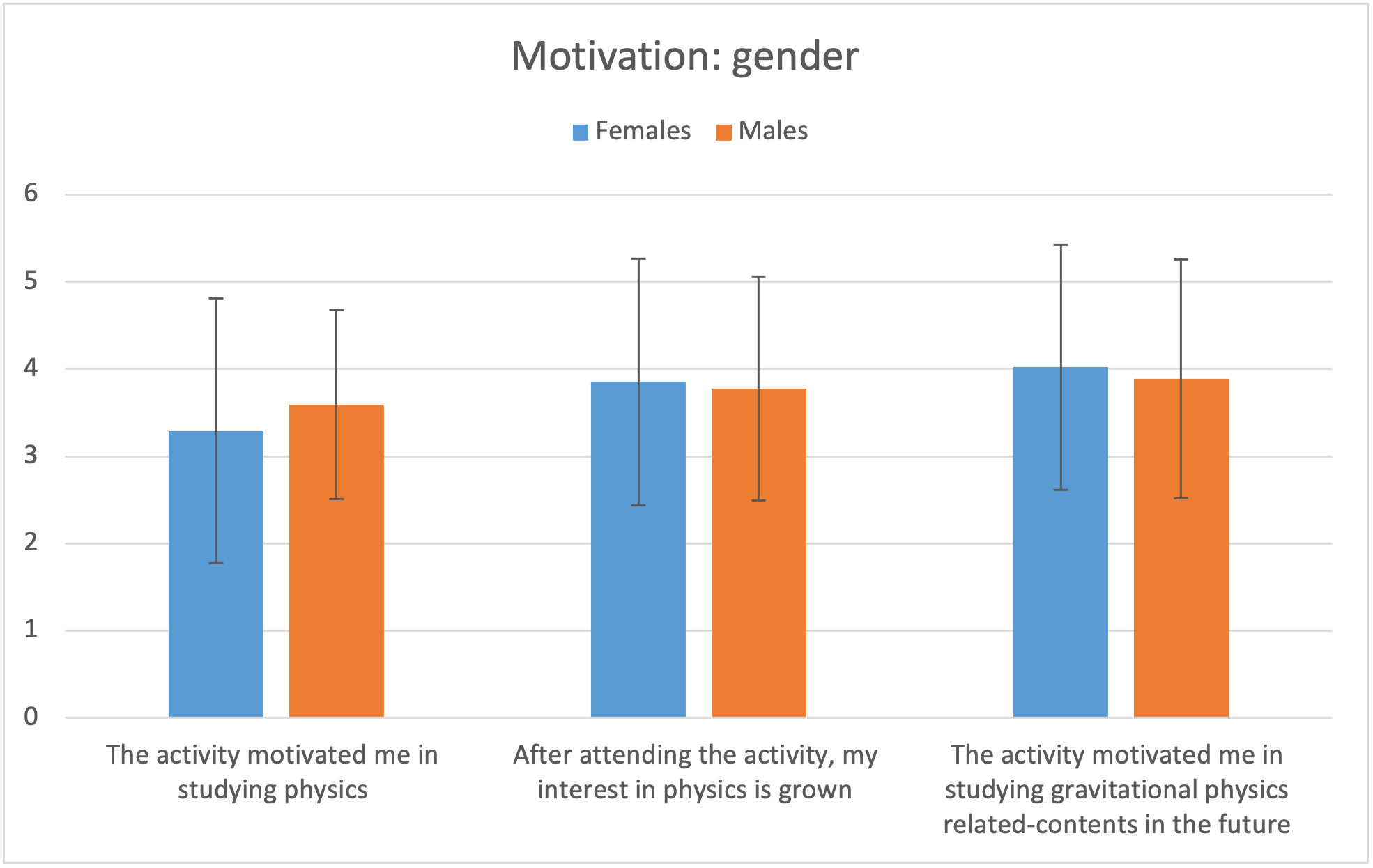}  
\caption{The panel shows means and standard deviation (error bars) about students' motivation in learning physics and the proposed subjects after attending the activity according to the gender.}\label{fig:motivation_gender}
\end{figure}

Concerning motivation, we investigated a) how the activity motivated students to study physics; b) if after attending the activity, their interest in physics grew; c) if the activity motivated them to study gravitational physics related-contents in their future.
Students rated items using a 6-point Likert scale, from 1 (completely disagree) to 6 (completely agree). The means and standard deviation for each item are shown in Fig.~\ref{fig:motivation}. We report results on single items according to the Likert scale in Table~\ref{table:motivation}. School, classes, and gender results are shown in Figs.~\ref{fig:motivation_school_classes} and~\ref{fig:motivation_gender}.

\begin{table}[h]\label{table:motivation}
\begin{tiny}
\begin{center}
\begin{tabular}{ | p{2.4cm}| | p{2cm}| p{1.8cm}| p{1.5cm}| p{1.8cm}| p{1.5cm}| p{1.8cm}| }
 \hline
 \multicolumn{7}{|c|}{Motivation in learning physics} \\
 \hline
  &  Completely Disagree & Disagree & Indifferent & Slightly Agree & Agree & Completely agree \\
 \hline
 The activity motivated me in studying physics   & 9 (11.8\%) & 9 (11.8\%)  & 20 (26.3\%)  & 22 (28.9\%) & 10 (13.2\%) & 6 (7.9\%)  \\
 & & & & & & \\
The activity motivated me in studying physics  & 3 (3.9\%) & 10 (13.2\%)  & 20 (26.3\%)  &  14 (18.4\%) & 20 (26.3\%) & 9 (11.8\%)  \\
& & & & & & \\
 The activity motivated me in studying gravitational physics related-contents in the future  & 4 (5.2\%) & 5 (7.9\%) & 18 (23.7\%) & 19 (25.0\%) & 16 (21.1\%) & 13 (17.1\%) \\
   \hline
  \end{tabular}
\end{center}
\caption{The table shows the number of answers (and corresponding percentage) for each point of the Likert scale concerning students' feedback about the use of poems as an educational tool.}
\end{tiny}
\end{table}

%FINAL REMARKS
Finally, when asked if they would participate again in similar activities, 66 students (86.8\%) answered \lq\lq yes\rq\rq, 3 (3.9\%) \lq\lq no\rq\rq, and 7 (9.2\%) \lq\lq I do not know\rq\rq. The overall feedback on the activity was as follows:  1 (1.3\%) rated as bad, 4 (5.3\%) were indifferent, 15 (19.7\%) slightly good, 30 (39.5\%) rated as good, and 26 (34.2\%) as very good. 

%%%

%%%%%%%%%%%%%%%%%%%%%%%%%%%%%%%%%%%%%%%%%%%%%%%%%%%%%%

%%%%%%%%%%%%
\subsection{Correlations and MANOVA}
Concerning correlations, results are reported in Table~\ref{table:correlations}. All of them are significative for p $<$ 0.1 (p stands for p-value).

The results of the MANOVA concerning differences between means of dimensions related to \lq\lq Preliminary knowledges\rq\rq, \lq\lq STEM Engagement\rq\rq, \lq\lq The Monologue\rq\rq, \lq\lq Poems\rq\rq, and \lq\lq Motivation\rq\rq~according to gender, school and classes showed a significant multivariate difference only for the school~(Wilkis lambda .704, F $=$ 5.05, p $<$ 0.1).In particular, the differences are associate to \lq\lq STEM Engagement\rq\rq, \lq\lq The Monologue\rq\rq, \lq\lq Poems\rq\rq. 
Concerning \lq\lq STEM Engagement\rq\rq, the means are higher for scientific schools, whereas, means are higher for humanities in the \lq\lq Monologue\lq\lq, and \lq\lq Poems\rq\rq domains, see Tables~\ref{table:STEM_Manova},~\ref{table:STEM_monologue}, ~\ref{table:STEM_Poem}.

\begin{table}[h]
\begin{tiny}
\begin{center}
\begin{tabular}{ | p{2cm}| p{2cm}| | p{2cm}| p{1.5cm}| p{1.8cm}| p{1.5cm}| p{1.8cm}| }
 \hline
 \multicolumn{7}{|c|}{Correlations} \\
 \hline
  &  & Preliminary Knowledges & STEM engagement & Monologue &Poems &Motivation \\
 \hline
 Preliminary Knowledges  &Pearson correlation & 1  & & & &  \\
   &  & & & &  & \\
 STEM engagement & Pearson correlation & .656  &  1  & & & \\
  &  &  & & & &  \\
 Monologue & Pearson correlation & .523 & .483  & 1 & & \\
  &  &  & & & & \\
 Poems  & Pearson correlation & .356  & .320   &  .734   & 1&  \\
  &  &  &  &  & & \\
 Motivation & Pearson correlation & .604 & .600  & .637 & .564 & 1  \\
 % &  &  &  & &  &  \\
  \hline
  \end{tabular}
\end{center}
\caption{The table shows correlations between macro-domains investigated. All of them are significative for p $<$ 0.1 . }
\label{table:correlations}
\end{tiny}
\end{table}

\begin{table}[h]
\begin{tiny}
\centering
\begin{tabular}{||c c c c c||} 
 \hline
 \multicolumn{5}{|c|}{STEM Engagement} \\
\hline
  & N & Mean & Std. Dev. & Std. Err. \\ [0.5ex] 
 \hline\hline
 Humanities & 23 & 4.26 & 0.92 & 0.19 \\ 
 Scientific & 52 & 4.54 & 0.79 & 0.11 \\
 Total & 75 & 4.46 & 0.84 & 0.09 \\  [1ex] 
 \hline
\end{tabular}
\caption{Means, standard deviation and standard error for the \lq\lq STEM Engagement\rq\rq domain}
\label{table:STEM_Manova}
\end{tiny}
\end{table}

\begin{table}[h]
\begin{tiny}
\centering
\begin{tabular}{||c c c c c||} 
 \hline
 \multicolumn{5}{|c|}{Monologue} \\
\hline
  & N & Mean & Std. Dev. & Std. Err. \\ [0.5ex] 
 \hline\hline
 Humanities & 23 & 4.48 & 0.96 & 0.20 \\ 
 Scientific & 52 & 4.11 & 0.97 & 0.13 \\
 Total & 75 & 4.23 & 0.97 & 0.11 \\  [1ex] 
 \hline
\end{tabular}
\caption{Means, standard deviation and standard error for the \lq\lq Monologue \rq\rq domain}
\label{table:STEM_monologue}
\end{tiny}
\end{table}

\begin{table}[h]
\begin{tiny}
\centering
\begin{tabular}{||c c c c c||} 
 \hline
 \multicolumn{5}{|c|}{Poems} \\
\hline
  & N & Mean & Std. Dev. & Std. Err. \\ [0.5ex] 
 \hline\hline
 Humanities & 23 & 4.41 & 0.99 & 0.21 \\ 
 Scientific & 52 & 3.68 & 1.21 & 0.17 \\
 Total & 75 & 3.91 & 1.19 & 0.14 \\  [1ex] 
 \hline
\end{tabular}
\caption{Means, standard deviation and standard error for the \lq\lq Poems \rq\rq domain}
\label{table:STEM_Poem}
\end{tiny}
\end{table}
%

%Finally, for each domain we investigated, we also made a T-test to see whether there were any statistically significant differences among the results we obtained according to schools, classes, and gender. There is only one case where significant differences are found, i.e. in the case of the use of poems as an educational tool. In particular, we found that means are higher for humanities than the scientific (two-variable T-test with unequal variances, two-tailed p $<$ 0.01), see also Fig.~\ref{fig:poems_school_classes} and Table~\ref{table:STEM_Poem}. 

%%%%%%%%%%%%%%%%%%%%%%%%%%%%%%%%%%%%%%%%%%%%%%%%%%%%%%%%%%
\section{Discussion}
We have studied the use of storytelling and artistic tools to bring contemporary physics topics and, in particular, black hole and gravitational waves physics at high school. The aim of this work was to explore the effect of storytelling on students' levels of classroom participation, motivation, and interest in the proposed topics. We also meant to measure their engagement and, most interestingly, their views about the effectiveness of storytelling as a teaching/learning strategy in the science classroom.
To do so, we designed a specific activity we proposed to 5 schools in Sardinia, Italy (200 students involved). The activity started with a monologue about the physics of ET and its possible implementation in Sardinia. The monologue made use of storytelling and poems, the latter being both in Italian and Sardinian language and related to the physical content of the story. After that, a 20-minute session to explain the physics of ET and of gravitational waves was done. In particular, we focused on gravity according to GR, black hole and gravitational wave formation, cosmology, and the early universe as possibly studied by ET. The activity ended with a 20-minute session of debate, where students asked questions and raised their curiosity about the proposed topics. 

To measure the efficacy of our activity we wrote a research questionnaire to investigate specific domains. What emerges from our results is quite encouraging in many aspects. 
Even if the majority of students did not know the ET project, infrastructure and which physics will be studied, regardless of the type of school or the class, they rated their average scientific preparation as adequate to attend the activity. Means are slightly higher for scientific and for the fifth class with respect to humanities and the third and fourth classes, respectively, but there are not any significative differences. This result is encouraging since it means that such topics can be discussed at school even without specific preparation for them. If we take a look at classes, students attending the third and fourth classes have a lower mean than their colleagues in the fifth classes. This is probably due to the fact that these contents are far away from the scholarly curriculum and that their preparation in mathematics and physics was at a starting level. Students did not know many concepts used during the monologue as the one of \lq\lq field\rq\rq~(e.g. electromagnetic field), or \lq\lq wave\rq\rq. Let us notice that during the monologue no use of mathematics was done.

In particular, students' higher interest was in black hole physics and cosmology. These findings are confirmed also in the case of the type of school, with means generally higher for humanities than the scientific. In the case of classes, the general trend is respected (black holes and cosmology are the preferred topics), and the means are higher for students attending the fifth class than the others. However, students in the third class showed higher means than their colleagues in the fourth class, even if there is not any significant difference between the two. Concerning gender, males showed slightly higher means than their female colleagues in all the items investigated within this domain (no significant differences). 

Students reported having a high interest in science communication, where we intended outreach and dissemination activity, see Fig.~\ref{fig:interest_outreach}. Also, in this case, there are not any significant differences between schools, classes, and gender (even if males exhibit a slightly higher mean on this item than females). Science communication as a tool to learn science was rated with high votes by most of the students, thus suggesting that activities like the one we designed can be considered as a supplementary tool in teaching and learning. One half of the sample rated with high votes the role of science communication in driving their interest in science and in orientating them to university (see Fig.~\ref{fig:stem_scales}). When asked if science communication activities motivated them to start an academic career, the mean raises down with respect to the other items. A possible interpretation of this result is that even if this kind of activity can, in some sense, help in orientating students toward a particular university course, raising their interest in science, it is not evident that they will pursue such a career. Nevertheless, students affirmed that through this kind of activity, they had entered into contact with science, learning new concepts, see Fig.~\ref{fig:monologue}.

The use of an artistic tool such as the monologue which, in turn, meant using storytelling to tell physics, passionated students towards the subject proposed. They also evaluated it as a useful instrument to communicate science. This suggests that implementing these strategies in the classroom during formal curriculum can engage students in science. Moreover, the majority of students affirmed to have learned new concepts of physics thanks to the monologue, raising their interest in physics, see Table~\ref{table:monologue}. Concerning creativity, we did not collect very high votes, but if we take a look at the type of schools, humanities showed higher means than scientific (even if there are not any significative differences between the two). If we want to understand better why the global mean is not so high, we can also take a look at the class distribution (see Fig.~\ref{fig:monologue_school_classes}). Students attending the third and fourth classes exhibited lower means than their colleagues attending the fifth class (and students attending the third and fourth classes represent the majority of the sample). This can be related to their lower level of preparation in physics, thus suggesting that the contents of the monologue could be more adequate for students attending the last two years of high school. However, if we take a look at the capacity of storytelling to passionate people, no differences between classes appeared in the data. Concerning gender, females reported slightly higher means than males (see Fig.~\ref{fig:monologue_gender}).

A similar situation appears in the use of poems within the monologue. What is really interesting is that when one takes a look at the school distribution of data, humanities reported higher means for every item we investigated. Results reveal that the difference is significant, thus suggesting that the use of artistic tools to communicate physics, as well as an educational tool, is welcome in humanities. This is an important result, also corroborated by the results in the other domains we investigated, since humanities suffer from the lack of science in their curriculum. Moreover, when we performed single informal interviews with a small random sample of students, the ones attending humanities reported that they are not good for pursuing a scientific career. If interdisciplinary approaches based on storytelling can serve in bringing science and, in particular, physics to school, teachers could think about the implementation of specific programs based on this methodology during their in-class activities. Students reported that the use of poems captured their attention, motivated them to attend the activity and, in the case of humanities, stimulated their creativity, too. The use of poems can offer students the possibility to put humanistic and scientific knowledge together, stimulating their critical thinking skills, as also suggested by~\cite{ref:Kragh2013} in the case of cosmology. This is also confirmed by the MANOVA (see Tables~\ref{table:STEM_Manova},~\ref{table:STEM_monologue},~\ref{table:STEM_Poem}), where significative differences according to the school appeared between the following domains: \lq\lq STEM Engagement\rq\rq, the \lq\lq Monologue\rq\rq,~and \lq\lq Poems\rq\rq. Students from humanities rated with higher votes the items appertaining to these domains, thus suggesting the positive influence of our methodology in engaging, interesting, and passionate them towards gravitational physics.  

Finally, concerning motivation, even if students seemed to be not very motivated in pursuing a career in physics (as also noted before in the case of science communication items), they want to learn more about gravitational waves physics. This, in some sense, seems to be counterintuitive. However, let us notice that for high school students, studying physics is different from studying astronomy or astrophysics, thus they can be interested in learning more gravitational physics but not in studying physics as a whole subject. This fact emerged during the debate phase of the activity, where students made questions to satisfy their need for knowledge about physics. Students made questions on black holes, white holes, dark matter and dark energy, the expansion of the universe, and quantum gravity. The expert stimulated the discussion, linking the answers to ET physics, whose fields of investigation will be very promising to address many of the students' questions~\cite{ref:Maggiore2020,ref:Branchesi2023}. The role of debate after science theatre has been emphasized in~\cite{ref:Giliberti2019,ref:Giliberti2022} as important to stimulate critical thinking and engage students in learning science, with particular attention to acquiring scientific literacy. Our results go in the same direction, which is to be taken into account by teachers during the design of a lesson.   

If we look at correlations between the domains we investigated, we see that \lq\lq Motivation\rq\rq correlates with the \lq Monologue\rq\rq and \lq\lq Poem\rq\rq domains. The correlation appears also with items related to \lq\lq STEM Engagement\rq\rq. This suggests that the efficacy of an activity like the one we proposed relies on students' sensibility towards science communication and its role in our society to learn physics. Nevertheless, since motivation correlates also with the methodology we used to bring physics into the classroom, a positive effect of storytelling and interdisciplinary in this domain appeared. This corroborates what is already known from literature on the positive effects of storytelling and interdisciplinary approaches to bring STEM in high school, fostering engagement and learning~\cite{ref:Rawatee2022,ref:Engel2018,ref:Kaur2020,ref:Simon2000,ref:Petrucco2009}. Studying correlations could offer some insights into the mechanism of generating enthusiasm and fostering motivation in students, possibly individuating the independent variables, which in our case are related to students' passion for science communication (outreach and dissemination). 

Finally, the overall feedback about the activity is very positive, showing students' appreciation of our methodology and its possible implementation at school. In particular, we also suggest that this kind of approach could be useful in orienting students toward a STEM or STEAM (Science Technology Engineering Arts, and Mathematics) career, as also discussed in~\cite{ref:Nguyen2018,ref:Devins2015}. In fact, what emerges from our results is that the activity raised students' motivation to learn gravitational-related topics. They also found poems and the monologue a useful tool to communicate physics, thus their interest can also be driven towards the study of arts and science communication (see also \cite{ref:Tuveri_Arts2022} and refs therein). Teachers in collaboration with researchers can design specific programs to integrate formal scholar curricula to engage students in these fields.

%%%%%%%%%%%%%%%%%%%%%%%%%%%%%%%%%%%%%%%%%%%%%%%%%%%%%%%%%%

\section{Conclusion}
In this work, we proposed storytelling and artistic tools (a theatrical monologue) to bring gravitational waves physics to high school. The PER group of the University of Cagliari and INFN Cagliari designed a specific program devoted to 200 high school students (17 to 19 years old, third to fifth classes) in Sardinia. 

The aim of our work was to explore the effect of storytelling on students' levels of classroom participation, motivation, and interest in the proposed topics. We also meant to measure their engagement and, most interestingly, their views about the effectiveness of storytelling as a teaching/learning strategy in the science classroom. We investigated these domains by self-report questionnaires administered to 76 students and data were analyzed both qualitatively and quantitatively. 

The results of our research are encouraging. We hope that they can also motivate high school teachers to explore this kind of approach in their classes, at least to introduce new arguments or to engage more students in STEM. Moreover, if teachers do not feel adequate to explore by themselves this methodology, they can promote students' creation of stories to motivate them to learn physics (see~\cite{ref:Rawatee2022,ref:Kotluk2016,ref:Marsico2019}). In particular, statistical analysis showed that results are more positive for humanities than scientific high school students. This is an important result since humanities suffer from the lack of a proper scientific curriculum. If interdisciplinary approaches based on storytelling can serve in bringing science and, in particular, physics to school, teachers could think about the implementation of specific programs based on a methodology like the one we proposed. 

Offering an interdisciplinary vision of science to high school students and teachers are becoming a common trend of informal learning (see~\cite{ref:Spelt2009,ref:Gao2020,ref:Davies2007} and refs therein). To show how science is evolving and to provide new instruments to learn science and physics in an enlarged context, mixing knowledge, techniques, and methods from different disciplines should be part of science and education curricula, developing an integrated model of learning and teaching~\cite{ref:Khalick2010}. 
This research can give instructors a methodological tool to encourage them to bring these topics to school, using storytelling to optimum advantage in science. It can be used to introduce students to current trends in research, trying to bypass content-related difficulties (both physical and mathematical), but still making them explore our universe with inquiry and minds-on activities, improving their motivation, curiosity, and interest in physics.

The study has some limitations that future investigations should take into account. The size of sample not allow to identify a model able to discriminate the cognitive factors important to improve efficacy in our methodology. Carry out interviews, also with a small sample, could help in this direction. A measure of the level of learning pre and post-activity (formative evaluation) could also give important feedback to quantitatively measure the efficacy of the methodology on learning.

%\begin{figure}
%\centering%
%\includegraphics[width=0.5\textwidth]{Template}  
%\caption{The panel shows the template to report students' results on their CPS activity within the Aria Project masterclass.}\label{fig:ariatemplate}
%\end{figure}

%%%%%%%%%%%%%%%%%%%%%%%%%%%%%%%%%%%%%%%%%%%%%

%%%%%%%%%%%%%%%%%%%%%%%%%%%%%%%%%%%
%%%%%%%%%%%%%%%%%%%%%%%%%%%%%%%%%%%

\section{Acknowledgments}
The authors acknowledge faculty members, teachers, and students who participated in our studies. 
%The authors also acknowledge the anonymous referees for the careful revision of the paper and for the helpful comments which raised the quality of the work. 

\section{Ethical statement}
Informed consent to participate in the study has been obtained from participants. Any identifiable individuals participating at the study have been also aware of intended publication. Informed consent to publish has be obtained from participants of the study. This work was carried out in accordance with the principles outlined in the journal's ethical policy and with the \lq\lq Codice etico e di comportamento\rq\rq~of the University of Cagliari.

\section{Funding declarations and conflicts of interests}
There are no known conflicts of interest associated with this publication and there has been no significant financial support for this work that could have influenced its outcome.

%%%%%%%%%%%%%%%%%%%%%%%%%%%%%%%%%%%%
%%%%%%%%%%%%%%%%%%%%%%%%%%%%%%%%%%%%

\end{document}